\documentclass[a4paper,12pt]{article}
\usepackage{amsmath}
\usepackage{amsfonts}
\usepackage{mathtools}
\usepackage{authblk}
\usepackage{graphicx,epsfig}
\usepackage{float}
\usepackage{subcaption}
\usepackage{amssymb}
\usepackage{physics}
\usepackage{cite}
\usepackage{mathtools}
\usepackage{soul}
\numberwithin{equation}{section}
\title{Photon region boundary for stationary axisymmetric black holes}
\author[1]{Prasad Padhye\footnote{ms18085@iisermohali.ac.in}\,}
\affil[1]{Indian Institute of Science Education \& Research (IISER), Mohali, India}
\author[2,3]{Kajol Paithankar\footnote{kajol.paithankar@iiap.res.in}\,}
\author[2]{Sanved Kolekar\footnote{sanved.kolekar@iiap.res.in} \,}
\affil[2]{Indian Institute of Astrophysics, Koramangala II Block, Bangalore 560034, India}
\affil[3]{Pondicherry University, R.V. Nagar, Kalapet 605014, Puducherry, India}

\begin{document}
\maketitle

\begin{abstract}

The black hole shadow is fundamentally connected to the structure of light rings and the photon region in the background geometry. We investigate the photon region boundary in a generic asymptotically flat, stationary, axisymmetric black hole spacetime that admits spherical photon orbits (SPOs). Explicit expressions possessing real solutions are provided for the photon region’s boundary purely in terms of the background metric functions, independent of the photon’s parameters like energy or angular momentum, which are applicable to both separable and non-separable spacetimes. We further analyze its common features, including overlap with the ergoregion and rotation sense of SPOs. Additionally, light rings are identified at the extrema of the photon region boundary curves in the $(r,\theta)$ plane. Our approach is validated against a few exact black hole solutions. Implications are discussed.

\end{abstract}

\section{Introduction}

The radio-imaging of the silhouette of the supermassive black holes M87* and Sgr A* released by the Event Horizon Telescope (EHT) collaboration \cite{EHT BH image M87, EHT BH image Milkyway} has opened up new avenues of probing strong field regimes and testing modified theories of gravity. Investigations of photon trajectories around black hole geometries and accretion physics play a key role in such studies. In the former, a comprehensive exploration of light rings, photon region, and their projection effects is significant in understanding how the black hole shadow appears to a distant observer for various black hole solutions in different gravity theories. For a review of such studies, see \cite{ShadhowBraneBH, ShadowRegularRotatingBH, ShadowAnalytic, ShadowEDGB_BH, FalckeBH, Volker Review, Horizon-scale tests of gravity theories, Constraining parameters, Rotating regular BHs, Shadows and strong gravitational lensing, Chaotic shadows of black holes, shadow radii}.

The black hole shadow and its boundary corresponding to the extent of the central dark region is essentially related to unstable null geodesics spiraling outwards from perturbed spherical photon orbits (SPOs). For spherically symmetric black holes, it is relatively straightforward to examine their photon spheres, whereas for rotating black holes, one needs to analyze both the light rings as well as the whole photon region wherein the SPOs are confined to lie. The photon region and its structure are crucial for determining the features of a black hole shadow. Though the structure of the shadow depends on the observer's position and motion in addition to the black hole's parameters, the photon region itself is characterized entirely by the black hole's geometry and depends only on the background spacetime metric.

The photon region and light rings (LRs) of the Kerr black hole have been elaborately described in the literature \cite{Carter, Exact formulas for SPOs in Kerr, ETeo Kerr2}. The two LRs outside the horizon, one counter-rotating and one co-rotating with the black hole, are situated at the intersection of the boundary of the photon region with the equatorial plane \cite{VolkerNotes}. For the Kerr-Newman-NUT black holes with a cosmological constant and the accelerated black holes of the Plebański–Demiański class, the analytical estimation of the photon region and the shadow can be found in \cite{PR of KN-NUT BH, PR of accelerated BH}. LRs have been extensively investigated in the spacetimes of exact black hole solutions as well as in generic backgrounds of spherically symmetric and stationary axisymmetric black holes and horizon-less ultra-compact objects. Based on the analysis of LRs around Kerr and Kerr-like black holes, the asymmetries of LRs of Kerr-like black holes have been obtained as a measure of violation of the no-hair theorem \cite{Photon Kerr}. In the backgrounds of rotating boson stars and Kerr black holes with scalar hair, the presence of a stable LR leads to trapped or quasi-bound orbits \cite{Chaotic lensing}. For a stationary axisymmetric horizon-less ultra-compact object, the LRs have been shown to occur in pairs, one at a saddle point and the other at a local extremum of the effective potential. One of the LRs is stable if the object is a solution of the Einstein field equations, which leads to nonlinear instabilities of the background spacetime \cite{LR Stability in UCOs, Long-lived modes}. If the ultra-compact object possesses an ergoregion, then at least one of its LRs lies outside the ergoregion \cite{LRs of stationary spacetimes}. In the case of stationary black holes, at least one LR occurs at a saddle point of the effective potential termed as a standard LR \cite{Stationary BHs and LRs}. Assuming the outer LR of the ultra-compact object to be standard, the inner LR has been proved to be stable, solely on the basis of geometric considerations \cite{Nature of inner LRs}. The outermost LR is radially unstable for non-extremal black holes, while for extremal black holes, at least one retrograde LR exists \cite{Universal properties of LR}. In quasi-black hole spacetimes, the presence of LRs is closely related to the instability of spacetimes\cite{quasiblack hole spacetimes}. The existence of LRs causes invariant structures in the phase space that are related to the black hole shadow \cite{Invariant structures}. 

Although the light rings have been well-established in stationary axisymmetric spacetimes, a similar thorough analytical inspection for the photon region is lacking. Photon regions have been mostly explored for exact black hole solutions with explicitly known metric functions. Given the exact black hole metric, the usual analytical method to derive the SPOs and the photon region in a $4$-dimensional separable stationary axisymmetric geometry is to identify the four constants of motion and then deduce the completely integrable form of null geodesic equations \cite{VolkerNotes}. In asymptotically flat spacetime, the conserved energy, angular momentum, and the Hamiltonian of a photon are three trivial constants, while the fourth constant is typically obtained through the separation of the radial ($r$) and polar ($\theta$) components of equations of motion using the Hamilton-Jacobi equation.

Furthermore, in general, a stationary axisymmetric spacetime may or may not admit SPOs. In the spacetimes that don't admit SPOs, one needs to consider photon orbits that are not confined to $r=$ constant hypersurfaces. One such spacetime is of the Kerr black hole with proca hair, wherein the null geodesic equations are not separable, and instead of SPOs, a more general class of photon orbits called the fundamental photon orbits (FPOs) exist. The existence of these FPOs further leads to an interesting feature of a cuspy edge in the black hole's shadow \cite{FPOs}. In general, it is not evident whether such FPOs lie in a closed region of spacetime and whether one can construct a photon region of such FPOs. Hence, in the current work, we restrict our investigations to the spacetimes that allow the existence of SPOs.

Assuming that the SPOs exist in a spacetime, the analytical conventional method cannot be implemented if the separability cannot be determined or a Carter-like constant is not known. Here, we provide an analytical technique for finding the photon region boundary, which is independent of the separation constant and based solely on the metric functions of the background spacetime. Our method thus can be applied to both separable and non-separable black hole spacetimes that admit SPOs. To our knowledge, such a technique does not exist in the literature yet, and our method is novel in this aspect.

In the non-separable cases, the conventional way is to evaluate the SPOs numerically using the coupled radial and polar components of equations on a case-by-case basis by choosing a set of initial conditions. However, this approach may be limited by the selection of initial conditions, and hence, the entire photon region, along with its true boundary, may not be recovered. Our work also helps to resolve this issue by providing the photon region's boundary as a set of initial conditions. With the expected location of the boundary known, one can numerically check whether SPOs and the photon region exist in the chosen background spacetime. This could lead to any of the two conclusions: that SPOs may or may not exist in the chosen spacetime.

We examine the photon region in a generic asymptotically flat, stationary, axisymmetric black hole spacetime admitting SPOs and having a Killing horizon. We write the background metric in the Boyer-Lindquist-type coordinates and obtain the photon region boundary considering the generic behavior of metric functions. Using the form of the metric functions at the Killing horizon and the asymptotic flat boundary, we show that the solution for the boundary exists in the domain of outer communication and further infer features of the photon regions that are common to such stationary axisymmetric black hole spacetimes. We follow the effective potential $V(r,\theta)$ approach adapted for probing the light rings around stationary axisymmetric ultra-compact objects \cite{LR Stability in UCOs}.

We first set up the background asymptotically flat, stationary, axisymmetric spacetime and discuss the constraint on the effective potential that needs to be satisfied at the location of SPOs in section \ref{Setup}. For the rest of the analysis, we assume that the spacetimes under consideration admit SPOs and the constraint is satisfied at the corresponding location. We then briefly describe the usual analytical method taken in the literature so far to analyze SPOs and photon region in separable spacetimes, and then introduce a generic approach towards photon region that applies to both separable and non-separable spacetimes in section \ref{Definition and existence}. The construction of the photon region's boundary based on the effective potential $V(r,\theta)$ is given in section \ref{Definition of boundary}. This construction depends on the photon's parameters, namely the energy $E$ and angular momentum $L$ of a photon. Since the photon region is expected to be characterized entirely by the black hole's geometry, we reformulate the boundary solely in terms of the metric functions of the background spacetime in section \ref{Inner and Outer boundary}. We then show in section \ref{Existence of boundary} that at least one solution for such a boundary lies outside the horizon. We further investigate salient features of these boundaries in section \ref{General features}. We analyze the intersection of the photon region with the ergoregion in section \ref{Intersection of the photon region and ergoregion}, the rotation sense of SPOs in section \ref{Rotation sense of SPOs}, and light rings in section \ref{Light Rings}. We show that the boundary of the photon region closer to the horizon called the inner boundary, may lie partially inside and partially outside the ergoregion, and the SPOs traversing it are co-rotating with the black hole. Whereas the boundary away from the horizon, termed the outer boundary, always lies outside the ergoregion, and the SPOs traversing the outer boundary are counter-rotating. Further, the light rings on the boundaries are located at the extrema of the curves $r(\theta)$ defining the boundaries of the photon region. As a consistency check, we implement our approach to known cases of exact black hole solutions for which the photon region has been studied using the conventional method in section \ref{Examples}. The discussion is presented in section \ref{conclusion}.

The metric signature adopted is $(-,+,+,+)$ and natural units $G=\hbar=c=1$ are used.

\section{Stationary axisymmetric black hole metric} \label{Setup}

We consider the most general $4$-dimensional stationary axisymmetric black hole spacetime in Boyer-Lindquist-type coordinates $(t,r,\theta,\phi)$ described by the metric \cite{Chandrasekhar},
\begin{eqnarray}
ds^2 &=& g_{tt} \,dt^2+ 2 g_{t\phi} \,dt\, d \phi +g_{\phi\phi} \,d \phi^2 +g_{rr} \,d r^2 +g_{\theta\theta} \, d\theta^2 \label{metric}
\end{eqnarray}
where, the metric coefficients $g_{ij}$ are smooth continuously differentiable functions of coordinates $r$ and $\theta$ and are independent of $t$ and $\phi$. The metric is thus adapted to the Killing vectors, $\partial_{t}$ and $\partial_{\phi}$, corresponding to the isometries of stationarity and axisymmetry. The convention for the rotation sense of the black hole is such that $g_{t\phi}<0$. Further, the black hole geometry has an event horizon at $r=r_H$ such that $r_H$ is the largest root of $g^{rr}(r,\theta)=0$ and a stationary limit surface (SLS) at $r=r_g$ such that $g_{tt}(r_g,\theta)=0$. Our region of interest is the domain of outer communication $r>r_H$. As per the signature adopted, the metric coefficients $g_{rr}$, $g_{\theta\theta}$ and $g_{\phi\phi}$ are positive everywhere outside the horizon, whereas the coefficient $g_{tt}$ is positive between the horizon and the SLS. Beyond the SLS, for $r>r_g$, we have $g_{tt}<0$. The ergoregion lies between the event horizon and the SLS, with $g_{tt}>0$ and $g_{rr}>0$ in this region. The spacetime considered is assumed to be asymptotically flat, thus reducing it to the standard Minkowski metric form in spherical coordinates $(t,r,\theta,\phi)$ at spatial infinity, with the metric functions simplifying to
\begin{equation}
    g_{tt} \to -1, \quad g_{t\phi} \to 0, \quad g_{\phi\phi} \to r^2 \sin^2\theta, \quad g_{rr} \to 1, \quad g_{\theta\theta} \to r^2
\end{equation}
Since we do not make any assumptions regarding the field equations, the results are valid in any metric theory of gravity wherein photons follow the null geodesics. In such black hole spacetimes, the Hamiltonian for null geodesics can be expressed as \cite{LR Stability in UCOs},
\begin{eqnarray}
H &=& \frac{1}{2} \left[ g^{rr}\,(p_r)^2 +g^{\theta\theta}\, (p_{\theta})^2 + g^{tt}\, (p_t)^2+ 2 g^{t\phi}\, (p_t\, p_\phi ) + g^{\phi \phi}\, (p_\phi)^2 \right]\nonumber\\
&=& \frac{1}{2} \left[K(r,\theta) + V(r,\theta)\right]\nonumber\\
&=& 0 \label{Hamiltonian}
\end{eqnarray}
where $p_i$ is the null four-momentum vector. The kinetic term $K(r,\theta)$ and the potential term $V(r,\theta)$ are defined as,
\begin{eqnarray}
    K(r,\theta) &=& g^{rr}\,(p_r)^2 +g^{\theta\theta}\, (p_{\theta})^2\label{Kinetic term}\\
    V(r,\theta) &=& g^{tt}\, (p_t)^2+ 2 g^{t\phi}\, (p_t\, p_\phi) + g^{\phi \phi}\, (p_\phi)^2\label{Potential term}
\end{eqnarray}
The potential $V(r,\theta)$ can also be expressed in terms of the covariant metric components $g_{tt}$, $g_{t\phi}$ and $g_{\phi \phi}$ using the conserved quantities derived from the symmetries of the spacetime. With the conserved quantities, namely the energy $E$ and the angular momentum $L$ of a photon, associated with the Killing vectors $\partial_{t}$ and $\partial_{\phi}$, the potential term takes the form \cite{LR Stability in UCOs}
\begin{eqnarray}
V &=& -\frac{1}{D}\,(g_{\phi\phi}\,E^2 +2\,g_{t\phi}\,E\,L+ g_{tt}\,L^2 ) \label{Potential}
\end{eqnarray}
where $D=g_{t\phi}^2-g_{tt} g_{\phi\phi}$ is the determinant of the 2-dimensional metric in $(t,\phi)$ spacetime. We consider the determinant $D$ to be non-zero everywhere outside the event horizon, thus enabling the existence of well-defined contravariant components of this 2-dimensional metric. Note that, outside the horizon, at the SLS, $g_{tt}=0$ and hence $D>0$. This further implies that the determinant $D$ is positive everywhere outside the event horizon. Furthermore, we assume the event horizon to be a Killing horizon with the metric functions following $g_{t\phi}^2-g_{tt} g_{\phi\phi}=0$ at the horizon $r_H$ \cite{Stationary BHs and LRs}. In the black hole spacetimes, we then have $D>0$ outside the event horizon and $D=0$ on the null hypersurface of the black hole horizon.

\subsection{Spherical Photon Orbits:}

In general, spherical photon orbits (SPOs) may or may not exist in the stationary axisymmetric spacetimes of the form of Eq.(\ref{metric}). A photon on a SPO needs to satisfy both $\dot{r} =0$ and $\ddot{r} = 0$ over the whole trajectory. One can use these conditions to obtain a constraint on the potential term, namely $\partial_r (g_{\theta\theta} V) =0$ as shown below. Then, an equivalent combination of say, $\dot{r} =0$ and $\partial_r (g_{\theta\theta} V) =0$ can be used to define SPOs. We briefly recall the derivation below, leading to this constraint on the potential term at the location of SPOs \cite{photon trapping orbits}.

For a photon moving on a constant $r$ hypersurface, $\dot{r}=0$ and $\ddot{r}=0$, which is equivalent to $p_r=0$ and $\dot{p_r}=0$. This simplifies Eq.(\ref{Hamiltonian}) to
\begin{eqnarray}
H &=& \frac{1}{2} \left[\,g^{\theta\theta}\, (p_{\theta})^2 + V(r,\theta) \,\right]=0 \label{Hamiltonian for SPO}
\end{eqnarray}
Further, the radial component of Hamilton's equation \cite{LR Stability in UCOs} simplifies to,
\begin{eqnarray}
\dot{p_r} &=& -\frac{1}{2}\left[\,(p_\theta)^2 \partial_r g^{\theta\theta} + \partial_r V(r,\theta)\,\right]=0 \label{radial Hamilton's equation for SPO}
\end{eqnarray}
Eliminating $p_\theta$ from Eqs. (\ref{Hamiltonian for SPO}) and (\ref{radial Hamilton's equation for SPO}), the constraint at the location of SPOs can be obtained to be,
\begin{eqnarray}
    \partial_r (g_{\theta\theta} V)&=& 0 \label{SPO condition}
\end{eqnarray}
One can further check that the above equation is consistent with the $\theta$ component of Hamilton's equations for SPOs. 

We can now redefine SPOs by replacing one of two conditions, $\dot{r}=0$ and $\ddot{r}=0$, with Eq.(\ref{SPO condition}). For the investigation of the photon region in the next section, we assume that the SPOs exist in the background black hole spacetime and the following two conditions are satisfied at the location of SPOs:
\begin{eqnarray}
    \partial_r (g_{\theta\theta} V)= 0 &\text{and}& \dot{r}=0 \label{final SPO conditions}
\end{eqnarray}

We would like to emphasize here that a stationary axisymmetric spacetime admitting SPOs may not necessarily be separable. It has been discussed in \cite{photon trapping orbits} that the existence of SPOs does not necessarily imply complete separability (The converse is always true). In particular, for a spacetime which admits SPOs, to have the separability property, the ratio of the metric components $g_{rr}$ and $g_{\theta\theta}$ needs to have a special form $g_{\theta\theta}/g_{rr}=f(r) h(\theta)$. This suggests that SPOs could potentially exist even in non-separable spacetimes. Since we are not considering any special form for the ratio $g_{\theta\theta}/g_{rr}$, our results are applicable to both separable and non-separable black hole spacetimes that admit SPOs.

In section \ref{Definition and existence}, we define the photon region boundary in terms of the potential $V(r,\theta)$, which makes it straightforward to check that Eq.(\ref{final SPO conditions}) is satisfied at the boundary. However, to establish the existence of a SPO, one must verify that Eq.(\ref{final SPO conditions}) is satisfied along the entire trajectory and not just at the turning points on the boundary. This ensures that the trajectory remains confined to a constant radius throughout its motion. We point out here that, in the current analysis, we are not verifying whether Eq.(\ref{final SPO conditions}) is valid inside the region enclosed by the photon region boundary, and hence we are not proving the existence of SPOs. Rather, we proceed under the assumption that SPOs do exist in the background spacetime and focus on the analytical study of the photon region boundary.

\section{Photon region and its boundary} \label{Definition and existence}

The standard definition of the photon region is the region in a black hole spacetime comprised of (stable or unstable) SPOs having a constant radial coordinate $r$ \cite{PR of KN-NUT BH, PR of accelerated BH, Invariant structures}. The set of photon geodesics parametrically very close to the unstable SPOs spiral outward, forming the image of the black hole shadow as seen by a distant observer. These SPOs may be located over a finite interval of radial coordinate $r$ given by, say, $r_{in}\leq r \leq r_{out}$ and each SPO, with a fixed $r$, may oscillate between the limiting values, say $\theta_{min}$ and $\theta_{max}$ of the polar coordinate $\theta$, depending on the explicit solution of the stationary axisymmetric black hole. A subset of these SPOs includes the orbits of the Killing vector $(\partial_t + \alpha\, \partial_\phi)$ called the light rings \cite{LR Stability in UCOs}. The photon region is reduced to a photon sphere in the case of non-rotating spherically symmetric black holes. In rotating black hole spacetimes, the co-rotating and counter-rotating photon orbits occur at different radii $r$ due to frame-dragging effects. This results in the formation of a photon region with an inner boundary close to the horizon given by, say, $r_i(\theta)$ and an outer boundary given by $r_o(\theta)$ with $r_i(\theta)<r_o(\theta)$ for all $\theta$.

We now briefly describe the conventional method followed in the literature to analytically probe the SPOs and the photon region, and then emphasize the necessity of a generalized approach. In a stationary black hole spacetime, the Hamiltonian $H=0$, the conserved energy $E$, and angular momentum $L$ of a photon are standard constants for the null geodesic motion. Incorporating these constants and the exact functional form of the metric components, the Hamilton-Jacobi equation, $H\left(x,\partial S/\partial x\right)=0$ (here $x$ stands for $(t,r,\theta,\phi)$) is solved assuming the ansatz $S=S_{t}(t)+S_{\phi}(\phi)+S_r(r)+S_{\theta}(\theta)$ \cite{VolkerNotes}. In separable spacetimes, this leads to the separation of radial and polar coupled equations with an additional constant that appears as the separation constant. In the Kerr spacetime, this separation constant is Carter's constant \cite{Carter}. In all the cases of known exact black hole solutions for which the SPOs and the black hole shadow have been explored analytically, a Carter-like constant has been identified, see \cite{PR of KN-NUT BH, PR of accelerated BH, plasma on Kerr spacetime, Spinning black holes, parametrized axially symmetric, BH in plasma and expanding universe, BH in presence of plasma, Superentropic black holes, JP BH} for examples. With the separate equations for $r$ and $\theta$ coordinates, the definition of SPOs is sufficient to locate the photon region. In non-separable spacetimes, wherein a Carter-like constant cannot be determined, one can still evaluate the SPOs numerically by solving the coupled radial and polar geodesic equations. For the numerical analysis, one needs to choose a set of initial conditions and solve the equations on a case-by-case basis. The recovery of the entire photon region, along with its true boundary, then significantly depends on the selection of appropriate initial conditions.

Consequently, our current understanding of the photon region is restricted to the exact known black hole solutions with the caveats highlighted above. We, therefore, need a technique that helps in examining common features of the photon region in a general stationary axisymmetric black hole spacetime admitting the SPOs. We illustrate such a technique that is independent of any constants of null geodesic motion and which utilizes only basic features of the metric functions describing the black holes' geometry, assuming the conditions in Eq.(\ref{final SPO conditions}) are satisfied at the location of SPOs. For numerical analysis, even without checking whether these conditions are satisfied or not, the photon region boundary established using this technique can further be utilized to specify the initial data. With the precise initial data, one can then numerically check whether SPOs and the photon region exist in the background spacetime under consideration.

In the following sections, we formulate the equations defining the boundary of the photon region and show that the solutions to these equations lie in the domain of outer communication for stationary axisymmetric black hole spacetimes described in section \ref{Setup}.

\subsection{Construction of Photon region} \label{Definition of boundary}

In a stationary axisymmetric spacetime defined by Eq.(\ref{metric}), the motion of a photon in the $(t,\phi)$ dimensions is governed by
\begin{eqnarray}
    \dot{t} &=& \frac{g_{t\phi} L+g_{\phi\phi} E}{D}\\
    \dot{\phi} &=&  -\frac{g_{t\phi} E+g_{tt} L}{D}
\end{eqnarray}
where $E$ and $L$ are the photon's energy and angular momentum, as measured by an asymptotic static observer, and the dot represents the derivative with respect to an affine parameter. Since all the metric components $g_{ij}$ depend on $r$ and $\theta$, the photon has only two degrees of freedom. The motion along $r$ and $\theta$ then forms the kinetic part of the Hamiltonian with $K(r,\theta) \geq 0$. The potential part then follows $V(r,\theta) \leq 0$ from Eq.(\ref{Hamiltonian}). This limits a photon's motion to the region in $(r,\theta)$ space where $V(r,\theta) \leq 0$. For each photon, in addition to the above equations in the $(t,\phi)$ dimensions, there exist $r$ and $\theta$ components of geodesic equations that are generally coupled equations.

We define the photon region as a collection of all spacetime events outside the event horizon such that at least one SPO passes through an event in the set. The SPOs comprising the photon region are null geodesics on a constant $r$ hypersurface, which by definition requires $\dot{r}=0$ and $\ddot{r}=0$. Further, a SPO in the photon region, though at fixed $r$, can have motion in the $\theta$ direction. We define the boundary of a photon region in the $(r, \theta)$ plane to be such that every point on the boundary is a turning point in $\theta$, that is $\dot{\theta}=0$, for at least one SPO passing through that point. The inner boundary of the photon region is then the collection of all such turning points $(r,\theta)$, which forms a continuous and smooth curve $r_i(\theta)$ of the least possible value of $r$ for each value of $\theta$. A similar description will hold for the outer boundary $r_o(\theta)$ but with the maximum possible value of $r$ for each $\theta$. By this definition, the photons will have $\dot{\theta}=0$ in addition to $\dot{r}=0$ and $\ddot{r}=0$ at the boundary of the photon region which is equivalent to $p_r=p_\theta=0$ along with $\dot{p_r}=0$. Using Eq.(\ref{Hamiltonian for SPO}) and (\ref{radial Hamilton's equation for SPO}), these three conditions can be reformulated into the following two constraints on the potential $V(r,\theta)$:
\begin{eqnarray}
    V(r,\theta) &=& 0 \label{Boundary condition 1}\\
    \partial_r V(r,\theta) &=& 0 \label{Boundary condition 2}
\end{eqnarray}
The boundary of the photon region is then a curve $r_b(\theta)$ which satisfies these two constraints simultaneously, and the photon region is simply the region enclosed by the curves $r_b(\theta)$. From the above constraint equations, it is evident that Eq.(\ref{final SPO conditions}) is satisfied at the photon region boundary. We would like to highlight that this construction of the photon region doesn't require a Carter-like constant at any step of the calculations and hence can be applied to non-separable black hole spacetimes. The above constraints on the potential imply that a photon traversing through a point, say $(r_b(\theta),\theta)$ on the photon region's boundary, encounters a potential $V(r_b,\theta)=0$ which is also a local radial extremum (minimum or maximum along r) of the potential, satisfying $\partial_r V(r,\theta)=0$ at $r=r_b$.

The light rings, as a subset of SPOs, solve an additional constraint equation and hence are expected to lie on the boundary of the photon region. Light rings are obtained as a solution say $\{r_L,\theta_L\}$, to the following constraints on the potential \cite{LR Stability in UCOs}:
\begin{eqnarray}
V(r,\theta) = 0, \;\;\; \partial_r V(r,\theta) = 0, \;\;\; \partial_\theta V(r,\theta) = 0 \label{LR constraints}
\end{eqnarray}
Thus, at the location of the light ring, the potential $V(r,\theta)$ is an extremum along both the coordinates $r$ and $\theta$. In the following subsection \ref{Light Rings}, we show that these light rings are located at the extrema of the boundary curves $r_i(\theta)$ and $r_o(\theta)$ of the inner and outer boundaries of the photon region.

\subsection{Inner and Outer boundary of the photon region}\label{Inner and Outer boundary}

The definition of photon region boundary given in terms of potential $V(r,\theta)$ has the shortcomings of depending on the photon's parameters $E$ and $L$. We now eliminate these two parameters using the two constraint equations, Eq.(\ref{Boundary condition 1}) and (\ref{Boundary condition 2}) and define the photon region's boundary completely in terms of the metric functions $g_{tt}$, $g_{t\phi}$ and $g_{\phi\phi}$. 

The potential $V(r,\theta)$ in Eq.(\ref{Potential}) can be re-expressed as,
\begin{eqnarray}
    V(r,\theta) &=& -\frac{L^2}{D}\,(g_{\phi\phi}\,\sigma^2 +2\,g_{t\phi}\,\sigma + g_{tt} ) \label{first condition}
\end{eqnarray}
where $\sigma= E/L$ and $D=g_{t\phi}^2-g_{tt} g_{\phi\phi}$. Since $D$ is non-zero outside the event horizon, a solution to the first constraint $V(r,\theta)=0$ can be obtained as roots of the quadratic polynomial in $\sigma$. The two roots $\sigma_{\pm}$ are given as
\begin{eqnarray}
\sigma_{\pm} &=& \frac{-g_{t\phi} \pm \sqrt{(g_{t\phi})^2 - g_{\phi\phi} g_{tt}}}{g_{\phi\phi}} \label{Sigma Solutions}
\end{eqnarray}
Considering the chosen convention for the rotation sense of the black hole, with $g_{t\phi}$ taken to be negative, we can comment on the nature of these roots inside and outside the ergoregion. Within the ergoregion, $g_{tt}>0$ implies that both $\sigma_+$ and $\sigma_-$ are positive, whereas outside the ergoregion, $g_{tt}<0$ results in a positive value for $\sigma_+$ and a negative value for $\sigma_-$. Here, one can note that, $\sigma_+ > \sigma_-$ regardless of the sign of $g_{t\phi}$. The roots $\sigma_{\pm}$ determine the rotation sense of photons on SPOs (see section \ref{Rotation sense of SPOs} for details). A photon with $\sigma>0$ $(\sigma<0)$ travels on a co-rotating (counter-rotating) orbit. The the second constraint $\partial_r V(r,\theta)=0$ can be expanded as follows,
\begin{equation}
     \partial_r V =  -\frac{L^2}{D}\,\left(\partial_r g_{\phi\phi}\,\sigma^2 +2\,\partial_r g_{t\phi}\,\sigma + \partial_r g_{tt} \right) - \left(g_{\phi\phi}\,\sigma^2 +2\,g_{t\phi}\,\sigma + g_{tt} \right)\,\partial_r \left( \frac{ L^2}{D}\right)=0\quad \label{second condition}
\end{equation}
The above equation is also a quadratic equation in $\sigma$ and can be solved for the two roots of the quadratic polynomial. However, considering that the photon region boundary is a simultaneous solution to the two constraints, Eq.(\ref{Boundary condition 1}) and (\ref{Boundary condition 2}), the above equation can be simplified further. Imposing the first constraint $V(r,\theta)=0$ results in vanishing of the second term in Eq.(\ref{second condition}), thus reducing it to,
\begin{eqnarray}
    (\partial_r g_{\phi\phi}\,\sigma^2 +2\,\partial_r g_{t\phi}\,\sigma + \partial_r g_{tt} ) &=& 0
\end{eqnarray}
Solving this equation for $\sigma$ we get two roots $\bar{\sigma}_\pm$ as
\begin{eqnarray}
\Bar{\sigma}_{\pm} &=& \frac{-\partial_rg_{t\phi} \pm \sqrt{(\partial_rg_{t\phi})^2 - \partial_r g_{\phi\phi} \ \partial_r g_{tt}}}{\partial_rg_{\phi\phi}} \label{Sigma Bar Solutions}
\end{eqnarray} 
Since $\Bar{\sigma}_{\pm}$ gives the ratio $E/L$ for a photon orbiting a SPO, we expect it to be finite except for the SPO near poles where $L\to 0$. The monotonic behavior of metric functions along the radial coordinate, in particular $\partial_r g_{\phi\phi}>0$ and $\partial_r g_{tt}<0$, ensures that the roots $\Bar{\sigma}_{\pm}$ are finite and real-valued. For further analysis, we consider that the metric functions of the black hole spacetimes follow $\partial_r g_{\phi\phi}>0$ and $\partial_r g_{tt}<0$. These conditions on the metric functions can be relaxed if one uses the $H_+$ and $H_-$ potentials as defined for light rings in \cite{LR Stability in UCOs} and discussed more in section \ref{Existence of boundary}.

The solutions $\Bar{\sigma}_+$ and $\Bar{\sigma}_-$ are positive and negative, respectively, everywhere outside the horizon. However, unlike the $\sigma_{\pm}$ solutions, one cannot comment on the rotation sense of photons based on the definition of the $\Bar{\sigma}_\pm$ solutions. Hence, we need to identify the correct pairs $\sigma_p=\Bar{\sigma}_q$ that give the inner and outer boundary of the photon region. To note the nature (positivity/negativity) of these solutions inside and outside the ergoregion, we tabulate these cases below in Table \ref{sigma solutions}.
\begin{table}[H]
\centering
\begin{tabular}{|c|c|c|}
\hline
 & \text{Inside ergoregion} & \text{Outside ergoregion}\\
 \hline
 $\sigma_+$ & positive & positive \\
 \hline
 $\sigma_-$ & positive & negative \\
 \hline
 $\Bar{\sigma}_+$ & positive & positive \\
 \hline
 $\Bar{\sigma}_-$ & negative & negative \\
 \hline
\end{tabular}
\caption{$\sigma$ and $\bar{\sigma}$ solutions inside and outside ergoregion}
\label{sigma solutions}
\end{table}

The SPO on the photon region boundary satisfies the two constraints, Eqs. (\ref{Boundary condition 1}) and (\ref{Boundary condition 2}) simultaneously. Hence for such a SPO we have $E/L=\sigma_p=\bar{\sigma}_q$, where $p$ and $q$ could be $+$ or $-$. There are four possible pairs of $\sigma_p$ and $\Bar {\sigma}_q$. Two of the four pairs define the inner and outer boundary of the photon region, while the other two are physically invalid. Here, we note that in an axisymmetric spacetime as $g_{t\phi}\neq 0$, the roots $\sigma_+$ and $\sigma_-$ can never be equal at any point $(r,\theta)$ outside the horizon. We now consider each $\Bar{\sigma}_q$ root and, based on simple rational arguments, pair it with the appropriate $\sigma_p$ root:

\begin{enumerate}
\item $\bar{\sigma}_-$: $\bar{\sigma}_-$ is negative everywhere outside the event horizon. Hence, a photon with $E/L=\Bar{\sigma}_-=\sigma_p$, moves on a counter-rotating orbit. Such an orbit can occur only outside the ergoregion, thus implying that $\Bar{\sigma}_-$ can only be paired with the $\sigma_-$ solution outside the ergoregion, and the pair gives a boundary of the photon region that lies outside an ergoregion. 

\item $\bar{\sigma}_+$: $\bar{\sigma}_+$ is positive outside the event horizon, and hence, a photon with $E/L=\bar{\sigma}_+=\sigma_p$, moves on a co-rotating orbit. Based on the nature of solutions in Table \ref{sigma solutions}, one can deduce that, the pair $\Bar{\sigma}_+=\sigma_-$ will give a curve $r(\theta)$ that is discontinuous on the stationary limit surface. As $\sigma_+$ is positive everywhere outside the event horizon, the second appropriate pair defining the photon region boundary is $\Bar{\sigma}_+=\sigma_+$.
\end{enumerate}

The curve $r(\theta)$ obtained from $\Bar{\sigma}_+=\sigma_+$ may lie partially inside and partially outside the ergoregion, whereas the one obtained from $\Bar{\sigma}_-=\sigma_-$ will lie completely outside the ergoregion. Thus, we can infer that $\bar{\sigma}_+=\sigma_+$ leads to the inner boundary $r_i(\theta)$, while $\bar{\sigma}_-=\sigma_-$ leads to the outer boundary $r_o(\theta)$ of the photon region. Here, we emphasize that the inner and outer boundaries are obtained as solutions to the equations that depend only on the metric functions of the background spacetime and are given as,
\begin{eqnarray}
    \frac{-g_{t\phi} + \sqrt{(g_{t\phi})^2 - g_{\phi\phi} g_{tt}}}{g_{\phi\phi}}\Bigg|_{r=r_i(\theta)} &=& \frac{-\partial_rg_{t\phi} + \sqrt{(\partial_rg_{t\phi})^2 - \partial_r g_{\phi\phi} \ \partial_r g_{tt}}}{\partial_rg_{\phi\phi}}\Bigg|_{r=r_i(\theta)}\quad \label{inner boundary equation}\\
    \frac{-g_{t\phi} - \sqrt{(g_{t\phi})^2 - g_{\phi\phi} g_{tt}}}{g_{\phi\phi}}\Bigg|_{r=r_o(\theta)} &=& \frac{-\partial_rg_{t\phi} - \sqrt{(\partial_rg_{t\phi})^2 - \partial_r g_{\phi\phi} \ \partial_r g_{tt}}}{\partial_rg_{\phi\phi}}\Bigg|_{r=r_o(\theta)}\quad\label{outer boundary equation}
\end{eqnarray}

In the next subsection, we check whether the solutions of the above proposed expressions exist outside the horizon.

\subsection{Existence of the photon region boundary} \label{Existence of boundary}

Following the definitions for inner and outer boundaries of the photon region in Eqs. (\ref{inner boundary equation}) and (\ref{outer boundary equation}), we now prove that the well-defined real solutions $r_{i}(\theta)$ and $r_{o}(\theta)$ exist in the domain of outer communication.

We first recall the approximations of metric functions near the Killing horizon given in \cite{Symmetries at stationary non static horizon}. The metric functions $g_{tt}$, $g_{t\phi}$ and $g_{\phi\phi}$ are related to the lapse function $N$ and the frame dragging velocity $\omega$ through the following relations:
\begin{eqnarray}
\omega =-\frac{g_{t\phi}}{g_{\phi\phi}} \qquad &\text{and}& \qquad N^2=\omega^2g_{\phi\phi}- g_{tt} \label{Lapse function and frame dragging velocity}
\end{eqnarray}
Based on the regularity of the the Ricci scalar $\mathcal{R}$ and the traceless part of the Ricci tensor squared, $\mathcal{R}_{\mu\nu}\mathcal{R}^{\mu\nu}-\frac{1}{4} \mathcal{R}^2$, at the horizon, the lapse function $N$, the frame-dragging velocity $\omega$ and the metric coefficient $g_{\phi\phi}$ can be approximated as,
\begin{eqnarray}
N(n,z) &\simeq & \kappa_H \, n +\mathcal{O}\left(n^3\right) \\
\omega(n,z) &\simeq & \omega_H+ \frac{\omega_2(z)\, n^2}{2}+ \mathcal{O}\left(n^3\right)\\  
g_{\phi\phi}(n,z) &\simeq & g_{\phi\phi}^H(z) +\frac{g_{\phi\phi}^{(2)}(z)\,n^2}{2} +\mathcal{O}\left(n^3\right) \label{series of g phiphi}
\end{eqnarray}
where $n$ represents the normal distance to the horizon and $z$ is the direction perpendicular to that of $n$. The constant $\kappa_H$ is the surface gravity and is non-negative on the horizon. Using these approximations and the relations in Eq.(\ref{Lapse function and frame dragging velocity}), we evaluate the roots $\sigma_{\pm}$ and $\Bar{\sigma}_\pm$ near the horizon. The leading-order terms for the roots are obtained as,
\begin{eqnarray}
\sigma_+ &=& \omega_H + \left(\frac{\kappa_H}{\sqrt{g_{\phi\phi}^H}}\right)\, n +\mathcal{O}\left(n^2\right)\\
\sigma_- &=& \omega_H - \left(\frac{\kappa_H}{\sqrt{g_{\phi\phi}^H}}\right)\, n +\mathcal{O}\left(n^2\right)\\
\bar{\sigma}_+ &=& \omega_H + \frac{g_{\phi\phi}^H}{g_{\phi\phi}^{(2)}}\,\omega_2 +\sqrt{\frac{2\left(\kappa_H\right)^2}{g_{\phi\phi}^{(2)}}+\left(\frac{g_{\phi\phi}^H}{g_{\phi\phi}^{(2)}}\,\omega_2\right)^2} +\mathcal{O}\left(n\right)\\
\bar{\sigma}_- &=& \omega_H + \frac{g_{\phi\phi}^H}{g_{\phi\phi}^{(2)}}\,\omega_2 -\sqrt{\frac{2\left(\kappa_H\right)^2}{g_{\phi\phi}^{(2)}}+\left(\frac{g_{\phi\phi}^H}{g_{\phi\phi}^{(2)}}\,\omega_2\right)^2} +\mathcal{O}\left(n\right)
\end{eqnarray}
Further, $g_{\phi\phi}>0$ and $\partial_r g_{\phi\phi}>0$ translate to $g_{\phi\phi}^H>0$ and $g_{\phi\phi}^{(2)}>0$, which imply
\begin{eqnarray}
   \left| \frac{g_{\phi\phi}^H}{g_{\phi\phi}^{(2)}}\omega_2 \right| &< & \sqrt{\frac{2\left(\kappa_H\right)^2}{g_{\phi\phi}^{(2)}}+\left(\frac{g_{\phi\phi}^H}{g_{\phi\phi}^{(2)}}\,\omega_2\right)^2}
\end{eqnarray}
Then in the limit $n\to 0$, we have,
\begin{eqnarray}
    \bar{\sigma}_+ > \omega_H &\text{and} & \bar{\sigma}_- < \omega_H
\end{eqnarray}
Thus, close to the horizon, the two pairs of roots satisfy
\begin{eqnarray}
   \bar{\sigma}_+ > \sigma_+ &\text{and} & \bar{\sigma}_- < \sigma_- \label{Near horizon relations}
\end{eqnarray}
We now find such inequalities between the two pairs near spatial infinity. As the spacetime is asymptotically flat, the metric functions near $r\to \infty$ can be approximated as \cite{Asymptotically flat spacetimes}
\begin{eqnarray}
 g_{tt} &\rightarrow& -1+ \frac{C_1(\theta)}{r} + \mathcal{O}\left(\frac{1}{r^2}\right) \\
 g_{t\phi} &\to &\frac{C_2(\theta)}{r}+ \mathcal{O}\left(\frac{1}{r^2}\right)\\
 g_{\phi\phi} &\to & r^2\sin^2{\theta}+\mathcal{O}\left(\frac{1}{r}\right)
\end{eqnarray}
Using these approximations in the definitions of $\sigma_{\pm}$ and $\Bar{\sigma}_\pm$, we get the asymptotic approximations of the four roots. The leading-order terms are given as,
\begin{eqnarray}
\sigma_{+} &=& \dfrac{1}{\sin\theta}\left( \frac{1}{r}\right) + \mathcal{O}\left(\frac{1}{r^3} \right)\\
\sigma_{-} &=& -\dfrac{1}{\sin\theta}\left( \frac{1}{r}\right) + \mathcal{O}\left(\frac{1}{r^3}\right)\\
\bar{\sigma}_+ &=& \dfrac{\sqrt{C_1(\theta)}}{\sqrt{2}\sin\theta}\left( \frac{1}{r^{3/2}}\right) + \mathcal{O}\left(\frac{1}{r^{5/2}}\right)\\
\bar{\sigma}_- &=& -\dfrac{\sqrt{C_1(\theta)}}{\sqrt{2}\sin\theta}\left( \frac{1}{r^{3/2}}\right) + \mathcal{O}\left(\frac{1}{r^{5/2}}\right)
\end{eqnarray}
Then, in $r\to \infty$ limit, we obtain the inequalities
\begin{eqnarray}
    \bar{\sigma}_+ < \sigma_+ &\text{and} &\bar{\sigma}_- > \sigma_- \label{Asymptotic relations}
\end{eqnarray}
A direct comparison between the relations in Eq.(\ref{Near horizon relations}) and (\ref{Asymptotic relations}) leads to the following conclusion: Between the black hole horizon and the spatial infinity, there exist at least two curves $r_i(\theta)$ and $r_o(\theta)$ at which $\Bar{\sigma}_+=\sigma_+$ and $\Bar{\sigma}_-=\sigma_-$ respectively. In other words, at least one inner boundary and one outer boundary of the photon region is present between the horizon and spatial infinity. Thus, the photon region is bounded in a stationary axisymmetric black hole spacetime that admits SPOs, described by the metric in section \ref{Setup}.

In the above discussion, we have assumed $\partial_r g_{\phi\phi}>0$ and $\partial_r g_{tt}<0$. However, these conditions on the metric functions can be relaxed if one uses the $H_+$ and $H_-$ potentials defined for light rings in \cite{LR Stability in UCOs}. In that case, the definition of the photon region boundary as given in Eqs. (\ref{Boundary condition 1}) and (\ref{Boundary condition 2}) in terms of the effective potential $V(r,\theta)$ translates to $\sigma=\sigma_+$ with $\partial_r \sigma_+ =0$ for the co-rotating SPO at the inner boundary and $\sigma=\sigma_-$ with $\partial_r \sigma_- =0$ for the counter-rotating SPO at the outer boundary. These will lead to the following relations on the metric functions
\begin{eqnarray}
    \frac{\left(g_{t\phi}-\sqrt{D}\right)\partial_r g_{\phi\phi}- g_{\phi\phi}\,\partial_rg_{t\phi}}{g_{\phi\phi}^2} +\frac{ \,2g_{t\phi}\,\partial_rg_{t\phi}-\partial_r \left(g_{tt} g_{\phi\phi}\right)\,}{2\sqrt{D}g_{\phi\phi}}  &=& 0\\
    \frac{\left(g_{t\phi}+\sqrt{D}\right)\partial_r g_{\phi\phi}- g_{\phi\phi}\,\partial_rg_{t\phi}}{g_{\phi\phi}^2} -\frac{ \,2g_{t\phi}\,\partial_rg_{t\phi}-\partial_r \left(g_{tt} g_{\phi\phi}\right)\,}{2\sqrt{D}g_{\phi\phi}} &=& 0
\end{eqnarray}
In the spacetimes wherein $\partial_r g_{\phi\phi}>0$ or $\partial_r g_{tt}<0$ is not satisfied outside the event horizon, one can use the above equations to obtain the photon region boundary. To prove the existence of a solution, a similar procedure can be followed by looking at the limits of the LHS in the above expressions near the horizon and at the spatial infinity to verify that they are of opposite signs and, hence, they are indeed equal to zero somewhere in between.  

\section{General features of the photon region}\label{General features}

Based on the definition of photon region given in section \ref{Inner and Outer boundary}, we now explore some features such as overlapping of ergoregion and photon region, the rotation sense of the SPOs, and light rings located on the photon region boundary. These features are common to the photon regions of the stationary axisymmetric asymptotically flat black hole spacetimes described in section \ref{Setup}.

\subsection{Overlapping of photon region and ergoregion}\label{Intersection of the photon region and ergoregion}

We first determine the position of the boundaries with respect to the ergoregion. As per the definitions of inner and outer boundaries in section \ref{Inner and Outer boundary}, the inner boundary of the photon region may lie partially inside and partially outside the ergoregion, whereas the outer boundary always lies outside the ergoregion. We revisit these arguments by examining the roots $\sigma_{\pm}$ and $\Bar{\sigma}_\pm$ at the stationary limit surface (SLS). At the SLS, $g_{tt}(r_g,\theta)=0$ which implies
\begin{eqnarray}
    \sigma_+ &=& \frac{-g_{t\phi} + |g_{t\phi}|}{g_{\phi\phi}}=\frac{2 |g_{t\phi}|}{g_{\phi\phi}}\\
    \sigma_- &=& \frac{-g_{t\phi} - |g_{t\phi}|}{g_{\phi\phi}}=0
\end{eqnarray}
where $g_{t\phi}=-|g_{t\phi}|$ as $g_{t\phi}$ is chosen to be negative. Since the root $\Bar{\sigma}_-$ is negative everywhere outside the event horizon (see table \ref{sigma solutions}), at the SLS it follows,
\begin{eqnarray}
    \bar{\sigma}_-|_{r=r_g} < \sigma_-|_{r=r_g}
\end{eqnarray}
Comparing this inequality with the one in Eq.(\ref{Asymptotic relations}), we conclude that the curve $r_o(\theta)$ at which $\Bar{\sigma}_-=\sigma_-$ lies between the SLS $(r=r_g)$ and spatial infinity $(r\to \infty)$. Thus, the outer boundary of the photon region lies outside the ergoregion. A similar discussion on the light rings in a stationary spacetime with an ergoregion can be found in \cite{LRs of stationary spacetimes}.

The inner boundary of the photon region is given by the curve $r_i(\theta)$ at which $\Bar{\sigma}_+=\sigma_+$. If $\Bar{\sigma}_+>\sigma_+$ at the SLS, then the asymptotic relation in Eq.(\ref{Asymptotic relations}) implies that the curve $r_i(\theta)$ lies between the SLS and spatial infinity, that is, the inner boundary lies outside the ergoregion.

If there exists a point, say $\left(r_g(\theta_{in}),\theta_{in}\right)$, where $\Bar{\sigma}_+=\sigma_+$ on the SLS, then the curve $r_i(\theta)$ intersects the SLS at this point and the inner boundary lies partially inside and partially outside the ergoregion. The point of intersection $\left(r_g(\theta_{in}),\theta_{in}\right)$ can be evaluated using following condition:
\begin{eqnarray}
\bar{\sigma}_+|_{r=r_g}=\sigma_+|_{r=r_g} &=& \frac{2 |g_{t\phi}|}{g_{\phi\phi}} \label{Intersection}
\end{eqnarray}

\subsection{Rotation sense of SPOs}\label{Rotation sense of SPOs}
We investigate the rotation sense of the photon orbits traversing the inner and outer boundaries of the photon region. The angular velocity of a photon with respect to a static asymptotic observer is given by the ratio $\Omega=\dot{\phi}/\dot{t}$. For a photon in the stationary axisymmetric set-up considered, we have
\begin{eqnarray}
    \dot{t} &=& \frac{g_{t\phi} L+g_{\phi\phi} E}{D}\label{Time variation}\\
    \dot{\phi} &=&  -\frac{g_{t\phi} E+g_{tt} L}{D}
\end{eqnarray}
which gives the angular velocity to be,
\begin{eqnarray}
    \Omega &=& -\frac{g_{t\phi} (E/L) +g_{tt}}{g_{t\phi} +g_{\phi\phi} (E/L)} \label{Rotation sense}
\end{eqnarray}
Further, the SPOs traversing the boundary of the photon region satisfy $V(r,\theta)=0$, which is equivalent to,
\begin{eqnarray}
    g_{\phi\phi}\,\sigma^2 +2\,g_{t\phi}\,\sigma + g_{tt} &=& 0
\end{eqnarray}
with $\sigma=E/L$. This equation can be rewritten as,
\begin{eqnarray}
    g_{t\phi}\, \sigma +g_{tt} &=& -\sigma \,\left(g_{\phi\phi}\,\sigma+ g_{t\phi}\right) 
\end{eqnarray}
Substituting this in Eq.(\ref{Rotation sense}) we get,
\begin{eqnarray}
    \Omega &=& \sigma
\end{eqnarray}
Thus, the rotation sense of the photon traversing the boundary of the photon region is determined by the ratio $\sigma=E/L$ of its conserved energy and angular momentum component. The Eq.(\ref{Time variation}), can be re-expressed as,
\begin{eqnarray}
    \frac{\dot{t}}{E} &=& \frac{g_{t\phi}+g_{\phi\phi}\sigma}{\sigma\; D}
\end{eqnarray}
On the inner boundary of the photon region, photons have $\sigma=\sigma_+$. Then using Eq.(\ref{Sigma Solutions}) we can write
\begin{eqnarray}
  \frac{\dot{t}}{E}\Bigg|_{\sigma=\sigma_+} &=& \frac{g_{t\phi}+g_{\phi\phi}\sigma_+}{\sigma_+ \; D}=\frac{+\sqrt{D}}{\sigma_+ \; D} =\frac{1}{\sigma_+ \; \sqrt{D}}
\end{eqnarray}
As $\sigma_+$ and $D$ are positive everywhere outside the horizon, the quantity $\dot{t}/E>0$ on the inner boundary. Since $\dot{t}>0$ is a necessary condition for a physical photon, we have $E>0$ on the inner boundary of the photon region. Similarly, using the definition of $\sigma_-$ we can show that,
\begin{eqnarray}
  \frac{\dot{t}}{E}\Bigg|_{\sigma=\sigma_-} &=& \frac{g_{t\phi}+g_{\phi\phi}\sigma_-}{\sigma_- \; D}=\frac{-\sqrt{D}}{\sigma_- \; D} =\frac{-1}{\sigma_- \; \sqrt{D}}  
\end{eqnarray}
As $\sigma_-$ is negative and $D$ is positive outside the ergoregion, the quantity $\dot{t}/E>0$ and hence $E>0$ on the outer boundary of the photon region.

Thus, from the viewpoint of an asymptotic observer, the photon trajectories traversing the outer boundary of the photon region have $\Omega=\sigma_-<0$, that is, $L<0$ and hence are counter-rotating. While the trajectories traversing the inner boundary with $\Omega=\sigma_+>0$ and hence $L>0$ are co-rotating. This further confirms that the outer boundary lies outside the ergoregion, as the counter-rotating trajectories can exist only outside the ergoregion. A similar analysis for the energy $E$ and the rotation velocity $\Omega$ of photons moving on light rings in the equatorial plane can be found in \cite{Chaotic lensing}.

\subsection{Light rings}\label{Light Rings}
We now show that the light rings are located at the extrema of the curves $r_i(\theta)$ and $r_o(\theta)$. Light rings are special cases of SPOs where the null geodesics are circular orbits confined to a fixed $\theta$. Such circular orbits exist where the boundary of the photon region is tangent to the $r$=constant hypersurface \cite{PR of KN-NUT BH}. Consider the coordinate plane $(r,\theta)$. The projection of $r$=constant hypersurface in this plane is a $r$=constant curve, and the projections of the photon region's boundaries are $r=r_i(\theta)$ and $r=r_o(\theta)$ curves. Since, at the position of light ring, say $(r_L,\theta_L)$, the boundary of the photon region and the $r$=constant hypersurface have a common tangent plane, the normal vectors $\eta_\mu$, to these two hypersurfaces when projected in the $(r,\theta)$ plane should coincide at $(r_L,\theta_L)$.

The $r$=constant hypersurface has only one non-vanishing component $\eta_r$ of the normal vector, whereas, for the boundaries of the photon region, one can evaluate two components $\eta_r$ and $\eta_\theta$. For the two normal vectors to coincide, the $\theta$ component of the normal, $\eta_\theta$, must vanish at the position of a light ring. The $\eta_\theta$ components for the curves of the inner and the outer boundaries of the photon region are found to be proportional to,
\begin{eqnarray}
    \eta_{\theta_i} &\propto& \frac{dr_i(\theta)}{d\theta} \quad \text{for inner boundary}\\
    \eta_{\theta_o} &\propto& \frac{dr_o(\theta)}{d\theta} \quad \text{for outer boundary}
\end{eqnarray}
Consequently, one can determine the coordinates of light rings as $\left(r_i(\theta_{L_i}),\theta_{L_i}\right)$ on the inner boundary and $\left(r_o(\theta_{L_o}),\theta_{L_o}\right)$ on the outer boundary, where $\theta_{L_i}$ and $\theta_{L_o}$ are evaluated using
\begin{eqnarray}
    \frac{dr_i(\theta)}{d\theta} \Big|_{\theta=\theta_{L_i}} =0 &\text{and}& \frac{dr_o(\theta)}{d\theta} \Big|_{\theta=\theta_{L_o}} =0 \, .
\end{eqnarray}
The light rings are thus located at the extrema of the curves $r_i(\theta)$ and $r_o(\theta)$. Since light rings satisfy Eqs. (\ref{LR constraints}) which involves one additional constraint $\partial_\theta V = 0$ compared to the definition of the photon region boundary, the extremum of $r_i(\theta)$ and $r_o(\theta)$ in the $\theta$ direction must be equivalent to $\ddot{\theta} = 0$ to maintain consistency with \cite{LR Stability in UCOs}.

\section{Implementation in different gravity theories}\label{Examples}

We now apply the definition of photon region and the features discussed in the previous sections to the cases of the Kerr and Kerr-Newman black hole solutions of Einstein's field equations and the Kerr-Sen black hole solution of heterotic string theory. These black hole solutions admit SPOs and satisfy the conditions $g_{\phi\phi}>0$, $\partial_r g_{tt}<0$ and $\partial_r g_{\phi\phi}>0$ outside the event horizon. For these black hole solutions, the photon region boundary has been evaluated using a Carter-like constant and separated radial and polar geodesic equations. Our technique to determine the photon region is shown to be consistent with these known solutions. We chose the well-studied Kerr, Kerr-Newman, and Kerr-Sen black hole spacetimes to demonstrate the consistency and correctness of our analytical expressions. In the following subsections, we construct the photon region boundary without identifying any Carter-like constant, thus illustrating our method.

\subsection{Kerr black hole}
\begin{figure}[H]
\centering
\begin{subfigure}{0.40\textwidth}
\centering
\includegraphics[scale=0.40]{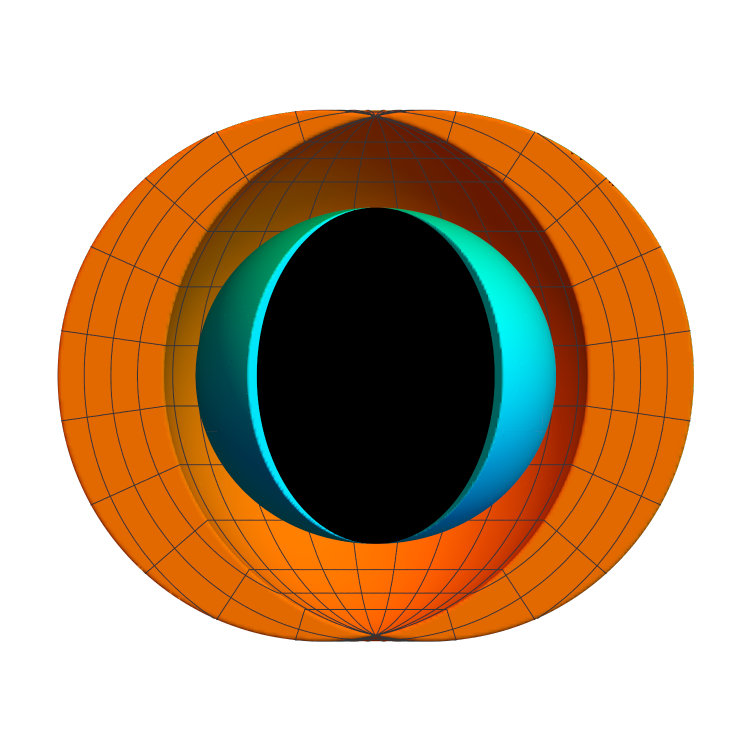}
\caption{$m=4$, $a=2$}
\end{subfigure}
\hfill
\begin{subfigure}{0.40\textwidth}
\centering
\includegraphics[scale=0.40]{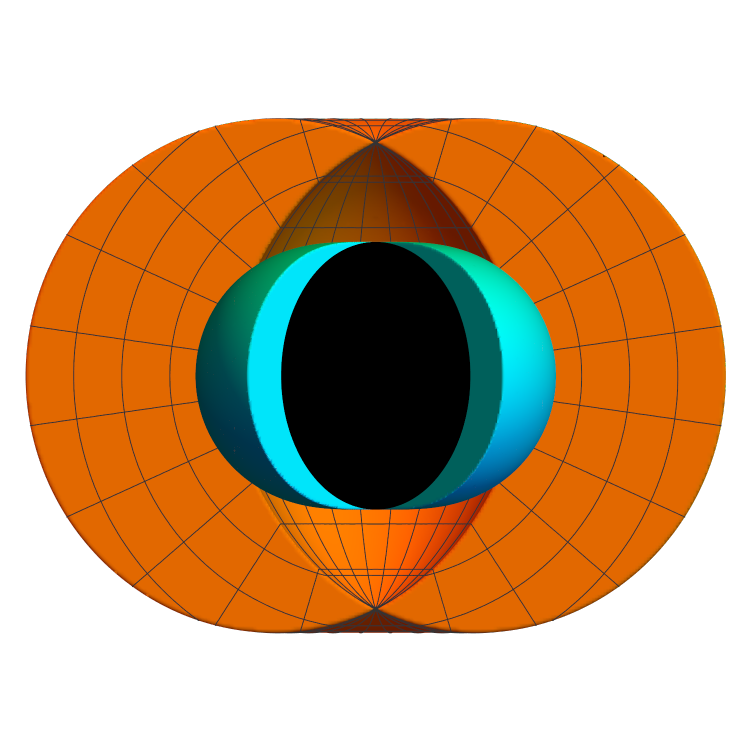}
\caption{$m=4$, $a=3.5$}
\end{subfigure}
\caption{3D illustrations showing the photon region in orange, ergoregion in cyan, and the event horizon in black for Kerr black holes.}
\label{Kerr illustration}
\end{figure}

\begin{figure}[H]
\centering
\begin{subfigure}{0.40\textwidth}
\centering
\includegraphics[scale=0.30]{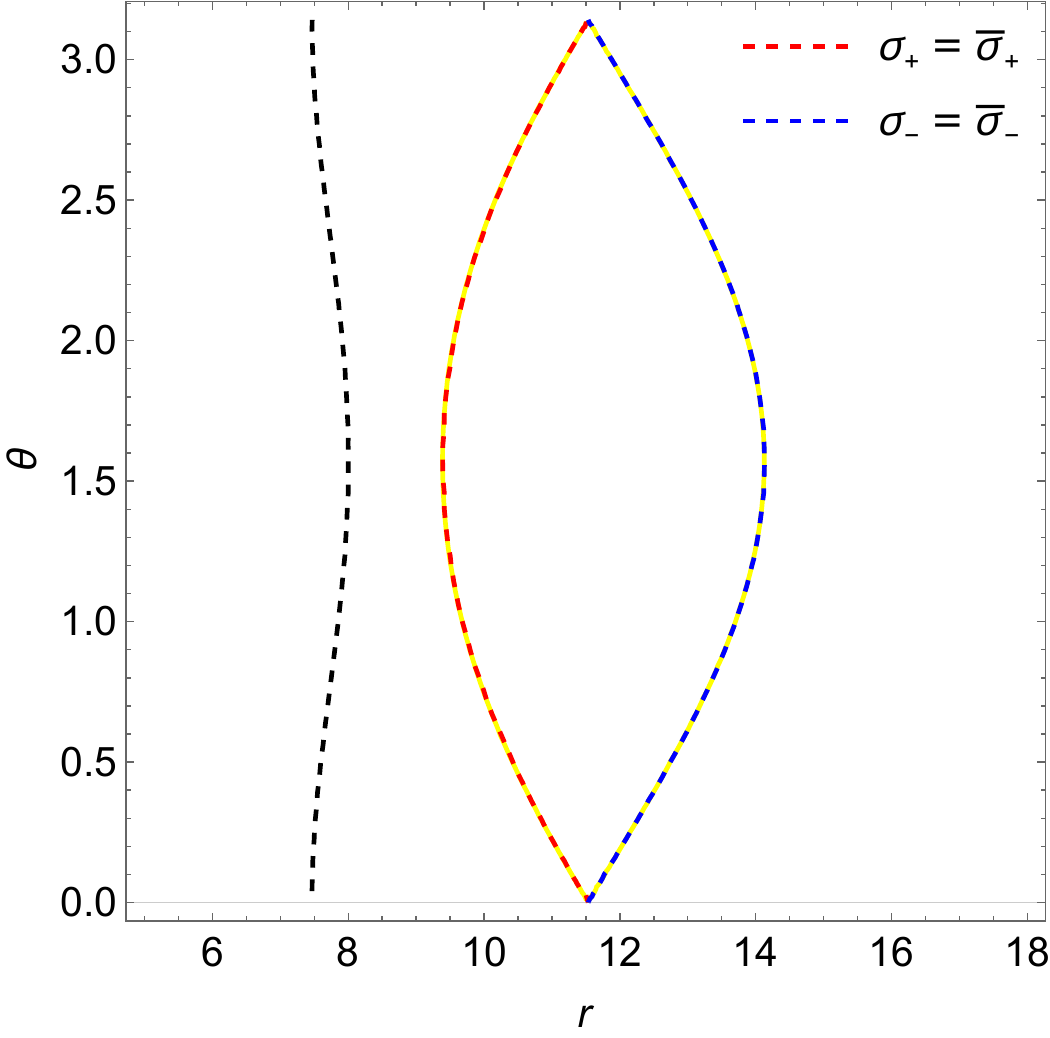}
\caption{$m=4$, $a=2$}
\end{subfigure}
\hfill
\begin{subfigure}{0.40\textwidth}
\centering
\includegraphics[scale=0.30]{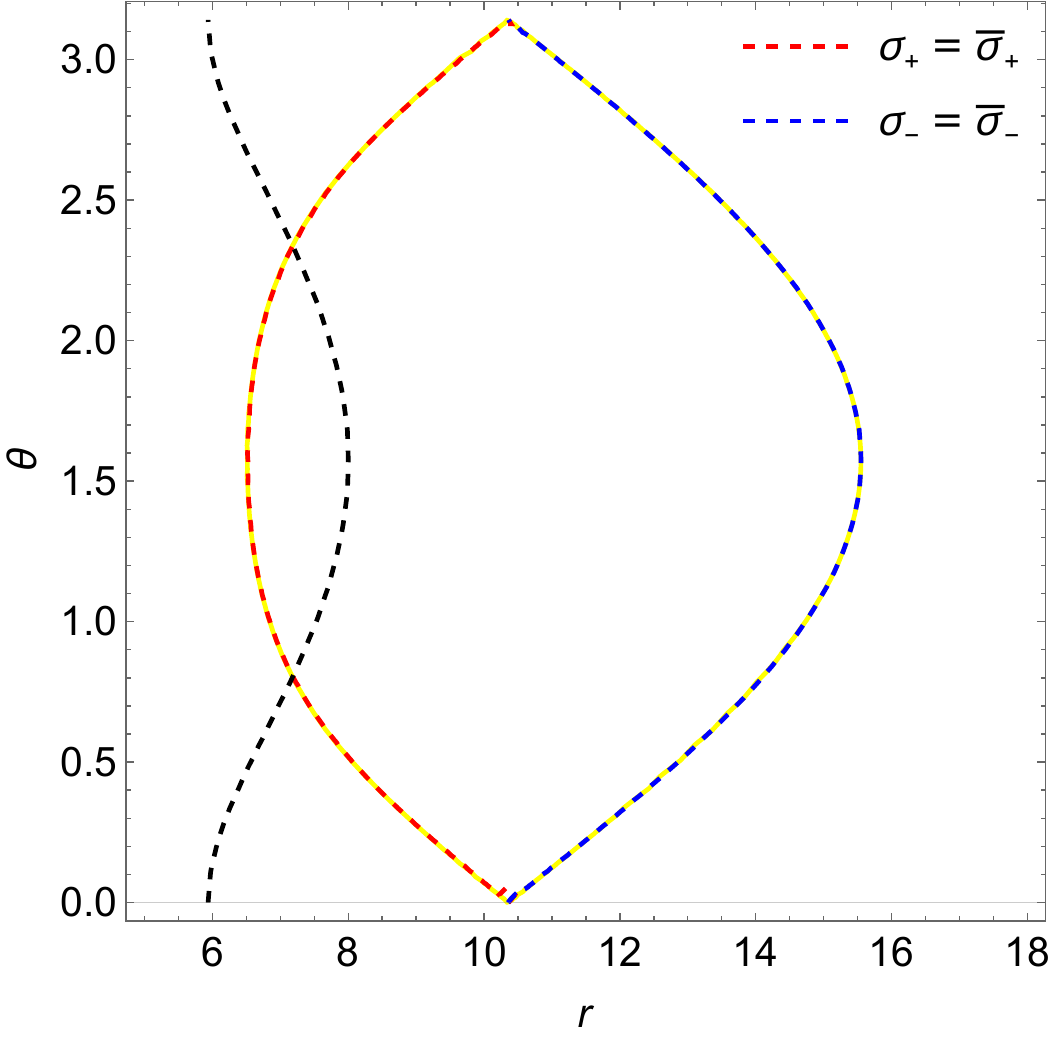}
\caption{$m=4$, $a=3.5$}
\end{subfigure}
\caption{The red and blue dashed lines show the curves $r_i(\theta)$ and $r_o(\theta)$ of the inner and outer boundaries. The yellow line represents the curve $r_c(\theta)$. The black dashed lines show the stationary limit surface $r_g(\theta)$.}
\label{consistency of Kerr boundaries}
\end{figure}

The metric for the Kerr black hole of mass $m$ and spin parameter $a$ in Boyer–Lindquist coordinates is given as,
\begin{eqnarray}
   ds^2 &=& -\left(1-\frac{2mr}{\rho^2}\right)\dd t^2 +\frac{\rho^2}{\Delta}\dd r^2+\rho^2\dd \theta^2 -\frac{4mra\sin^2{\theta}}{\rho^2}\dd t \, \dd \phi \notag \\
   & &{+\sin^2{\theta}\left(r^2+a^2+\frac{2mra^2\sin^2{\theta}}{\rho^2}\right)\dd \phi^2}
\end{eqnarray}
where $\Delta=r^2 +a^2 -2mr$ and $\rho^2=r^2 +a^2\cos^2{\theta}$. The metric represents the Kerr black hole for $a\leq m$. The radii of the horizon $r_H$ and the stationary limit surface $r_g$ are given by,
\begin{eqnarray}
    r_H &=& m+ \sqrt{m^2-a^2} \\
    r_g &=& m+\sqrt{m^2-a^2\cos^2\theta}
\end{eqnarray}
The separated equations for $r$ and $\theta$ coordinates with Carter's constant result in an inequality representing the photon region, which is given as \cite{VolkerNotes},
\begin{eqnarray}
\left(\rho^2(r-m)-2r\Delta \right)^2 &\leq & 4r^2\Delta a^2 \sin^2\theta \label{Kerr boundary}
\end{eqnarray}
Using this inequality, 3D illustrations of the photon region along with the black hole horizon, and the ergoregion are plotted in Fig. \ref{Kerr illustration} for the two sets of black hole parameters $\{m, a\}=(4,2)$ and $(4,3.5)$. 
Now, using the Eqs. (\ref{inner boundary equation}) and (\ref{outer boundary equation}), the curves $r_i(\theta)$ and $r_o(\theta)$ corresponding to the inner and outer boundaries are plotted along with the stationary limit surface at $r_g(\theta)$ in Fig. \ref{consistency of Kerr boundaries}. For comparison, the boundary curves $r_c(\theta)$ determined through the conventional method are also included in the same figure. The curves $r_c(\theta)$ satisfy
\begin{eqnarray}
\left(\rho^2(r-m)-2r\Delta \right)^2 &=& 4r^2\Delta a^2 \sin^2\theta
\end{eqnarray}
The overlapping of curves $r_i(\theta)$ and $r_o(\theta)$ with the curves $r_c(\theta)$ validates our technique in Kerr black hole spacetime. Further, the extrema of these curves lie in the equatorial plane, thus implying the existence of co-rotating and counter-rotating equatorial light rings \cite{VolkerNotes}.

\begin{figure}[H]
    \centering
    \includegraphics[scale=0.30]{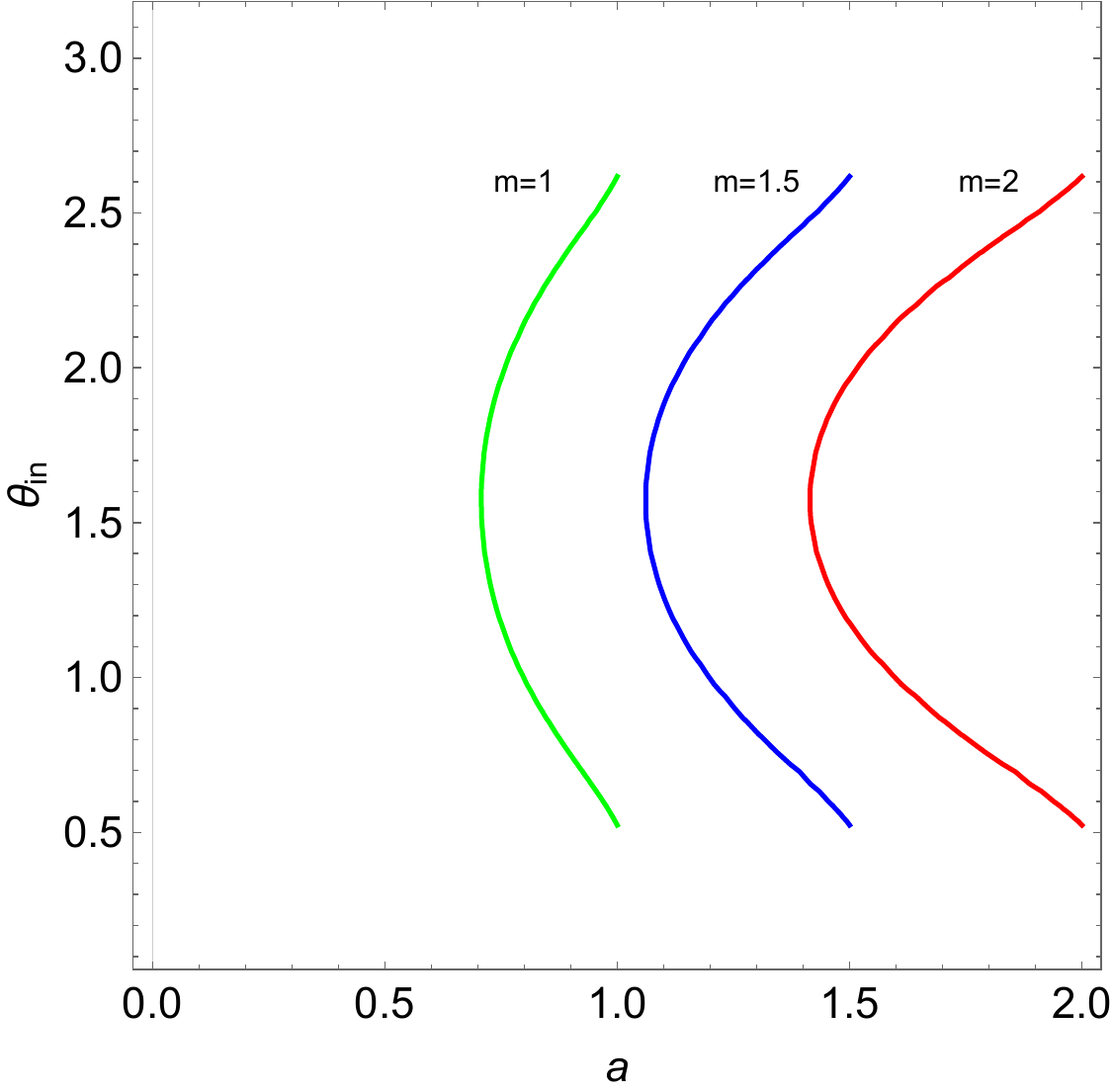}
    \caption{Intersection points $\theta_{in}$ for black hole mass $m=1,1.5$ and $2$ and the spin parameter $a<m$}
    \label{variation with spin in Kerr}
\end{figure}

The intersection of the inner boundary with the stationary limit surface is found by solving Eq.(\ref{Intersection}) for the intersection points $\{r_g(\theta_{in}),\theta_{in}\}$. For a fixed value of mass parameter $m$, the intersection points differ over the range of the spin parameter $a\leq m$. The intersection points $\theta_{in}$ are plotted over the range $a\leq m$ of the spin parameters in figure \ref{variation with spin in Kerr} below for $m=\{1,1.5,2\}$.
From the figure, one can note that corresponding to each mass parameter $m$, there exists a critical spin value, say $a_c$, beyond which the inner boundary of the photon region doesn't intersect the stationary limit surface. Consequently, for a Kerr black hole of spin $a<a_c$, the inner boundary of the photon region lies completely outside the ergoregion. For $a>a_c$, there exist two intersection points $\theta_{in}$ for each spin $a$, whereas at $a=a_c$, only one intersection point $\theta_{in}$ occurs at the equatorial plane $\theta=\pi/2$. The critical spin $a_c$ can be expressed as a function of mass $m$ of the black hole and is given by \cite{PR of KN-NUT BH},
\begin{eqnarray}
    a_c &=& \frac{m}{\sqrt{2}}
\end{eqnarray}
In the following subsections, we conduct a similar analysis for the Kerr-Newman and Kerr-Sen black holes. We find that, as in the case of the Kerr black hole, the boundary of the photon region obtained using the separated geodesic equations coincides with the one calculated using our technique for both black holes. Further, a critical spin value exists in both cases, beyond which the inner boundary of the photon region does not intersect the ergoregion and lies completely outside the ergoregion. The plots showing the intersection points and the expressions for critical spin parameters are presented for each of the black hole solutions in the corresponding subsections.

\subsection{Kerr-Newman Black Hole}

\begin{figure}[H]
\centering
\begin{subfigure}{0.40\textwidth}
\centering
\includegraphics[scale=0.40]{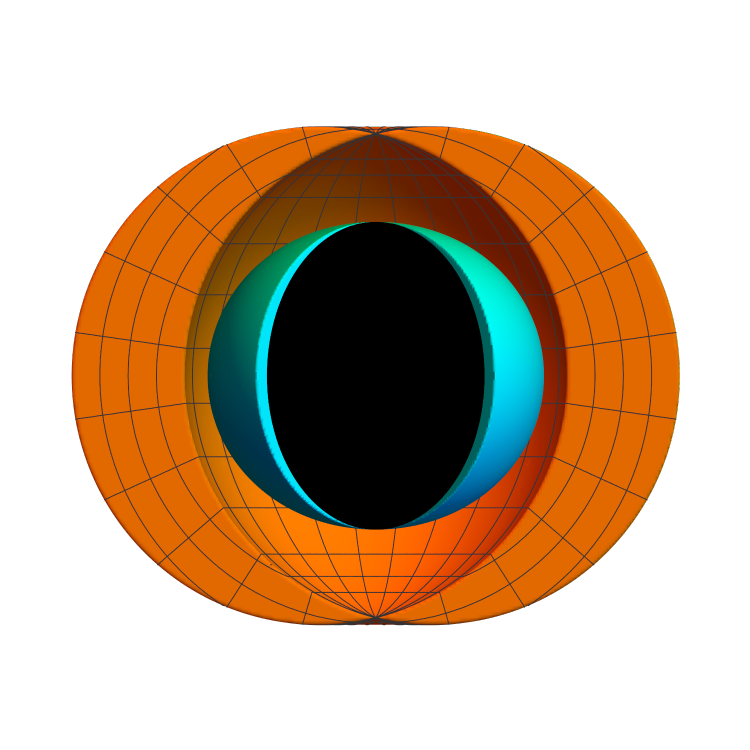}
\caption{$M=4$, $a=2$, $Q=2$}
\end{subfigure}
\hfill
\begin{subfigure}{0.40\textwidth}
\centering
\includegraphics[scale=0.40]{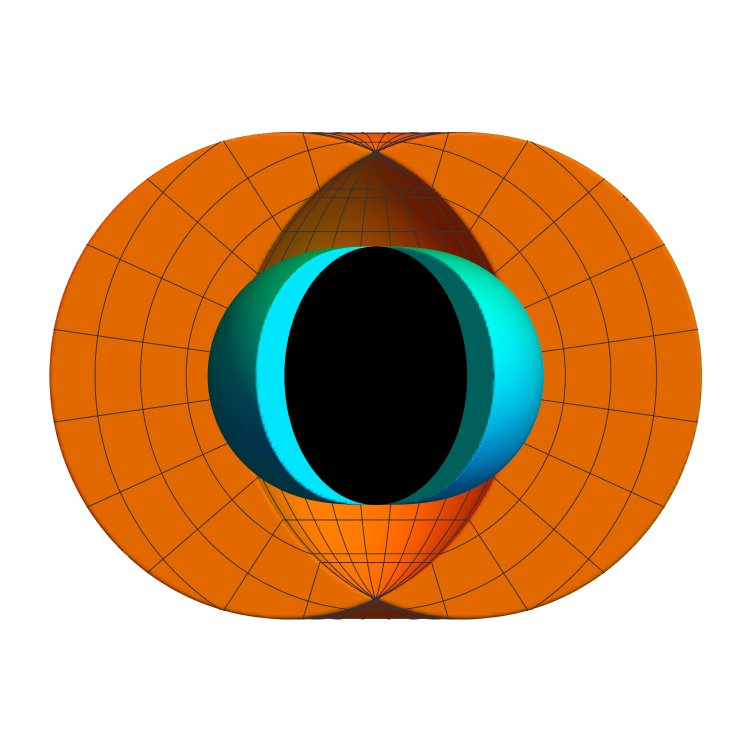}
\caption{$M=4$, $a=3$, $Q=2$}
\end{subfigure}
\caption{3D illustrations showing the photon region in orange, ergoregion in cyan, and the event horizon in black for Kerr-Newman black holes.}
\label{Kerr-Newman illustration}
\end{figure}

\begin{figure}[H]
\centering
\begin{subfigure}{0.40\textwidth}
\centering
\includegraphics[scale=0.30]{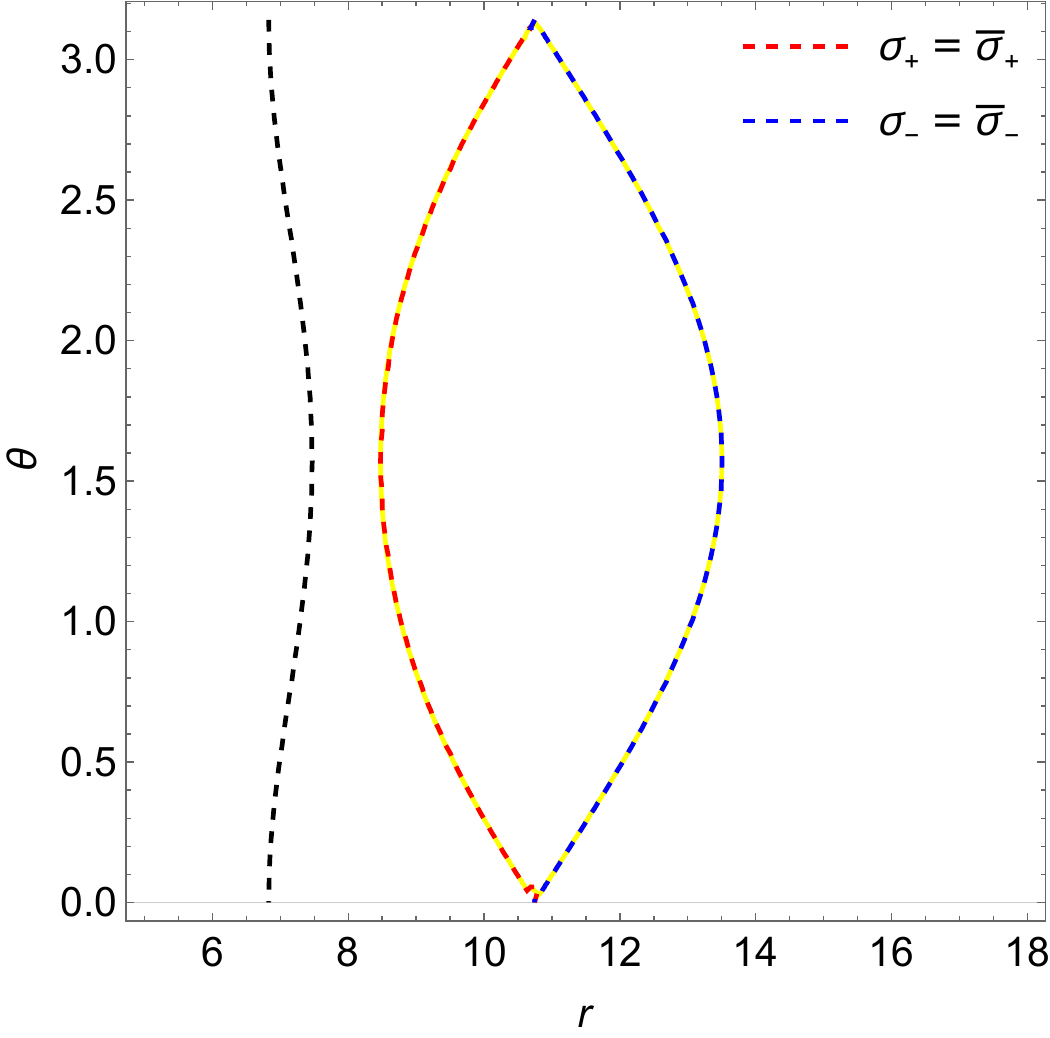}
\caption{$M=4$, $a=2$, $Q=2$}
\end{subfigure}
\hfill
\begin{subfigure}{0.40\textwidth}
\centering
\includegraphics[scale=0.30]{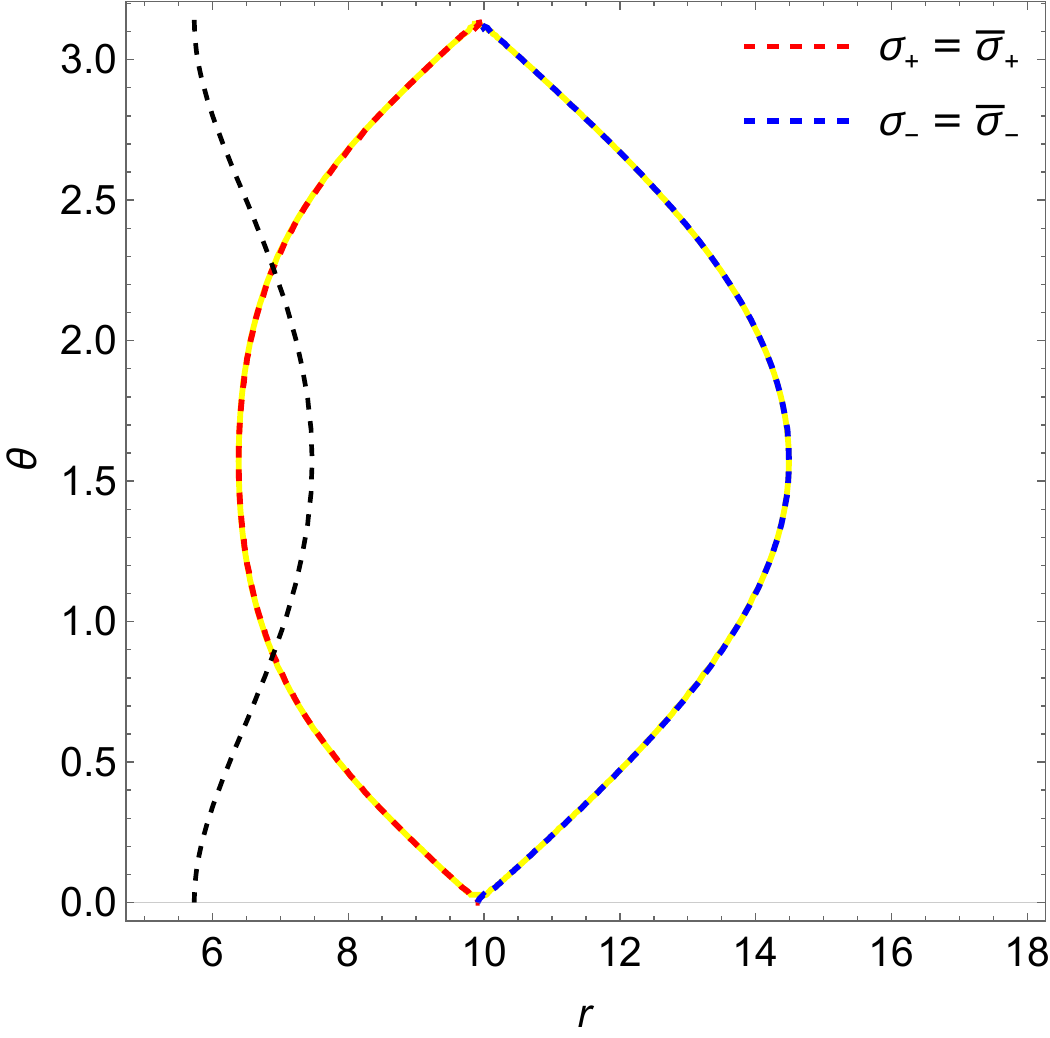}
\caption{$M=4$, $a=3$, $Q=2$}
\end{subfigure}
\caption{The red and blue dashed lines show the curves $r_i(\theta)$ and $r_o(\theta)$ of the inner and outer boundaries. The yellow line represents the curve $r_c(\theta)$. The black dashed lines show the stationary limit surface $r_g(\theta)$.}
\label{consistency of Kerr-Newman boundaries}
\end{figure}

The Kerr-Newman black hole of mass $M$, spin $a$ and charge $Q$ in Boyer-Lindquist coordinates is given by \cite{KN and KS BHs}
\begin{eqnarray}
    ds^2 &=& -\left(1-\frac{2Mr-Q^2}{\rho^2}\right)\dd t^2 +\frac{\rho^2}{\Delta}\dd r^2+\rho^2\dd \theta^2 -\frac{4Mra\sin^2{\theta}-2aQ^2\sin^2{\theta}}{\rho^2}\dd t \, \dd \phi \nonumber \\
   & & + \sin^2{\theta}\left(\frac{(r^2+a^2)^2-\Delta a^2\sin^2{\theta}}{\rho^2}\right)\dd \phi^2
\end{eqnarray}
where $\Delta=r^2-2Mr+a^2+Q^2$ and $\rho^2=r^2+a^2\cos^2\theta$ and $M^2\geq (Q^2+a^2)$. The event horizon and the stationary limit surface are located at $r_H$ and $r_g$ and are given by
\begin{eqnarray}
    r_H &=& M +\sqrt{M^2-Q^2-a^2}\\
    r_g(\theta)&=& M+\sqrt{M^2-Q^2-a^2 \cos^2\theta}
\end{eqnarray}
Using the separated radial and polar geodesic equations in \cite{KN and KS BHs}, the inequality describing the photon region can be evaluated to be,
\begin{eqnarray}
   \left( \eta_{KN}+a^2 \cos^2\theta\right) \sin^2\theta &\geq& \Phi_{KN}^2 \cos^2\theta
\end{eqnarray}
where
\begin{eqnarray}
    \eta_{KN} &=& -\frac{r^2\left[4a^2\left(Q^2-Mr\right)+\left(r^2-3Mr+2Q^2\right)^2\right]}{a^2(M-r)^2}\nonumber\\
    \Phi_{KN} &=& \frac{a^2(M-r)+r\left(r^2-3Mr+2Q^2\right)}{a(M-r)}\nonumber
\end{eqnarray}
The photon region, ergoregion, and the black hole horizon are plotted in Fig. \ref{Kerr-Newman illustration} for black hole parameters $\{M, a, Q\}=\{4,2,2\}$ and $\{4,3,2\}$. Then, using the Eqs. (\ref{inner boundary equation}) and (\ref{outer boundary equation}), the inner boundary $r_i(\theta)$ and outer boundary $r_o(\theta)$ are plotted along with the stationary limit surface in Fig. \ref{consistency of Kerr-Newman boundaries}. Additionally, the boundary $r_c(\theta)$ obtained using the separated radial and polar geodesic equations is also plotted in the same figure, which solves
\begin{eqnarray}
    \left( \eta_{KN}+a^2 \cos^2\theta\right) \sin^2\theta &=& \Phi_{KN}^2 \cos^2\theta
\end{eqnarray}

The curves $r_c(\theta)$ coincide with $r_i(\theta)$ and $r_o(\theta)$, which shows that our approach is consistent with the conventional method in the case of the Kerr-Newman black hole.

\begin{figure}[H]
\centering
\begin{subfigure}{0.40\textwidth}
\centering
\includegraphics[scale=0.30]{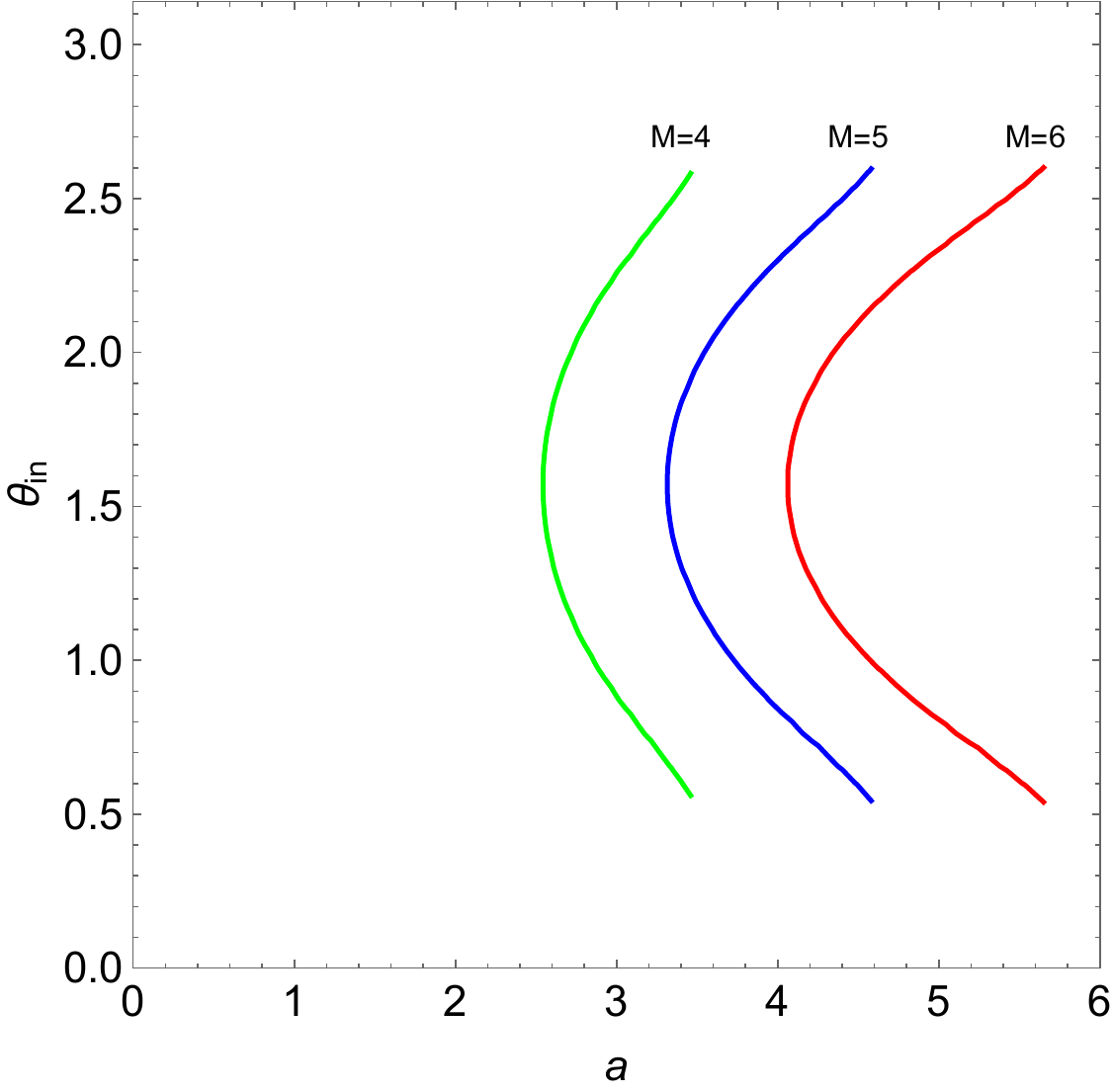}
\caption{$Q=2$}
\end{subfigure}
\hfill
\begin{subfigure}{0.40\textwidth}
\centering
\includegraphics[scale=0.30]{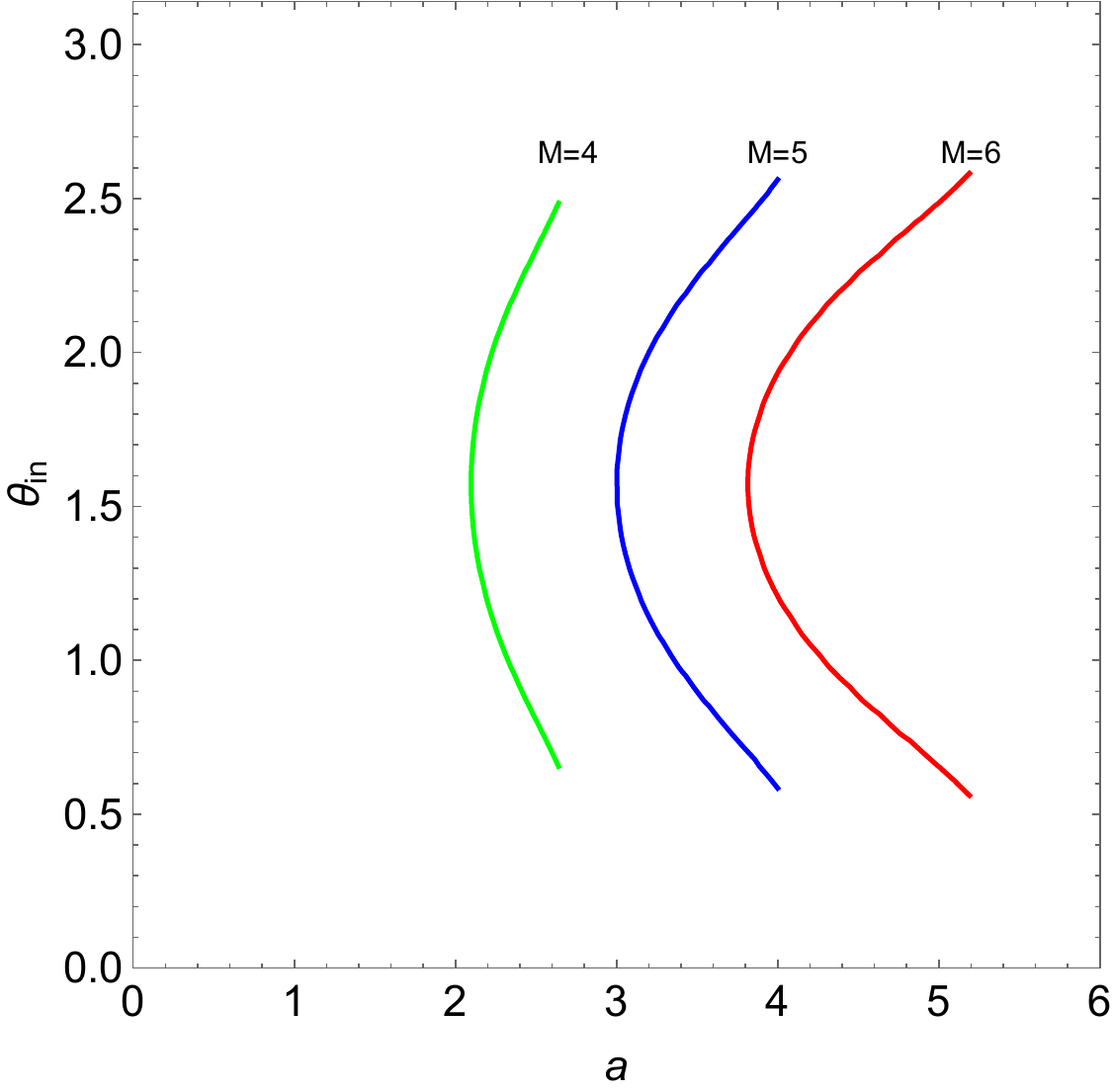}
\caption{$Q=3$}
\end{subfigure}
\caption{Intersection points $\theta_{in}$ for black hole mass $M=4,5,6$ and the spin parameter $a<\sqrt{M^2-Q^2}$}\label{variation with spin in Kerr-Newman}
\end{figure}

The intersection points $\theta_{in}$ of the inner boundary and the stationary limit surface are evaluated as solutions to the Eq.(\ref{Intersection}) and are plotted against the spin parameter over the range $a<\sqrt{M^2-Q^2}$ for black holes of mass $M=\{4,5,6\}$ and charge $Q=\{2,3\}$ in Fig. \ref{variation with spin in Kerr-Newman}.
The critical spin at which the inner boundary intersects the stationary limit surface at the equatorial plane $\theta=\pi/2$ is derived as a function of mass $M$ and charge $Q$ of the black hole and is given by
\begin{eqnarray}
a_c &=& \frac{\mathcal{A}_{M,Q}+\sqrt{2 M\, R_g\, \mathcal{B}_{M,Q}\, - Q^2 \,\mathcal{H}_M(Q)\,\mathcal{H}_M(-Q)}}{4 \left(Q^2 - 2 M R_g\right)\sqrt{ \mathcal{F}_{M,Q}}}
\end{eqnarray}
where
\begin{eqnarray} 
\mathcal{A}_{M,Q} &=& Q^2 \left(40 M^4-32 M^2 Q^2+3 Q^4\right)-M R_g\,\left(80 M^4-84 M^2 Q^2+17Q^4\right)\nonumber\\
\mathcal{B}_{M,Q}&=& \left(48 M^4-44 M^2 Q^2+7 Q^4\right)\left(48 M^6-68 M^4 Q^2+23 M^2 Q^4-Q^6\right)\nonumber\\
\mathcal{F}_{M,Q} &=& M R_g \left(16 M^4-20 M^2 Q^2+5 Q^4\right)-Q^2 \left(8 M^4-8 M^2 Q^2+Q^4\right)\nonumber\\
\mathcal{H}_M(Q) &=& 48 M^5-24 M^4 Q-44 M^3 Q^2+16 M^2 Q^3+7 M Q^4-Q^5\nonumber
\end{eqnarray}
with $R_g=r_g(\theta=\pi/2)= M+\sqrt{M^2-Q^2}$.

\subsection{Kerr-Sen Black Hole}
\begin{figure}[H]
\centering
\begin{subfigure}{0.40\textwidth}
\centering
\includegraphics[scale=0.30]{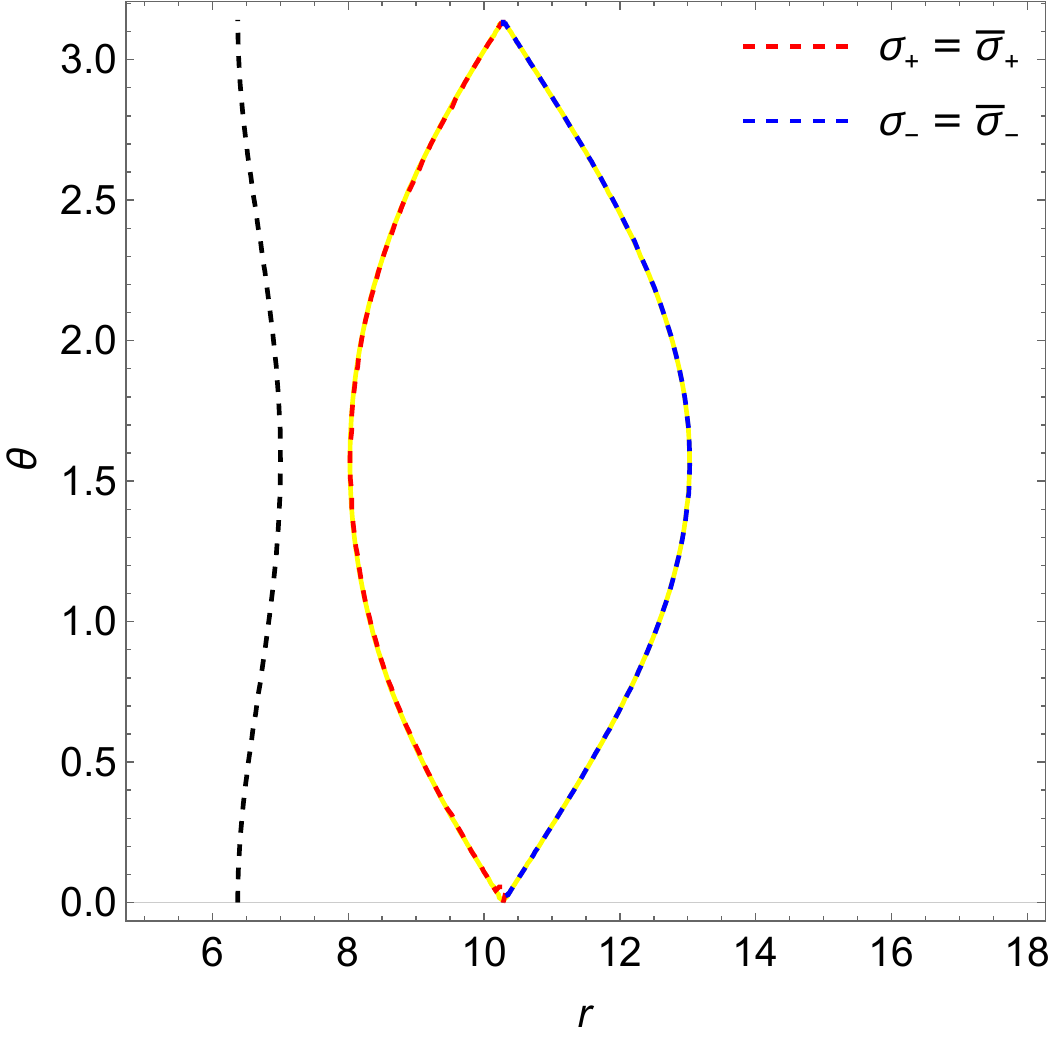}
\caption{$M=4$, $a=2$, $Q=2$}
\end{subfigure}
\hfill
\begin{subfigure}{0.40\textwidth}
\centering
\includegraphics[scale=0.30]{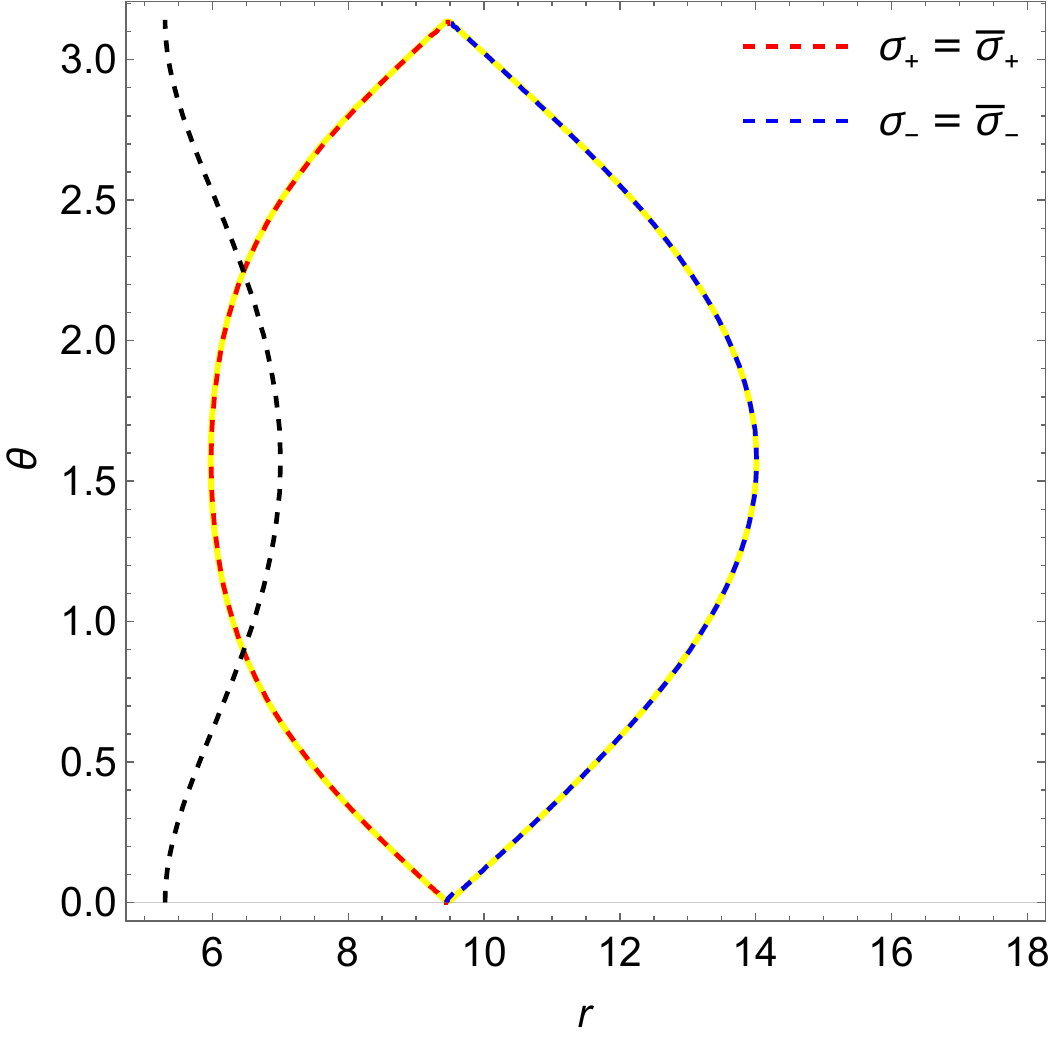}
\caption{$m=4$, $a=3$, $Q=2$}
\end{subfigure}
\caption{The red and blue dashed lines show the curves $r_i(\theta)$ and $r_o(\theta)$ of the inner and outer boundaries. The yellow line represents the curve $r_c(\theta)$. The black dashed lines show the stationary limit surface $r_g(\theta)$.}
\label{consistency of Kerr-Sen boundaries}
\end{figure}

The line element of the Kerr-Sen black hole of mass $M$, spin $a$ and charge $Q$ in Boyer-Lindquist coordinates is \cite{KN and KS BHs,KS BH}
\begin{eqnarray}
    ds^2 &=& -\left(1-\frac{2Mr}{\rho^2}\right)dt^2 + \rho^2 \left(\frac{dr^2}{\Delta}+d\theta^2\right)-\frac{4Mra\sin^2\theta}{\rho^2}dt d\phi\nonumber\\
    & & +\sin^2\theta \left(r^2+a^2+\frac{Q^2 r}{M}+\frac{2Mra^2\sin^2\theta}{\rho^2}\right) d\phi^2
\end{eqnarray}
where
\begin{eqnarray}
    \Delta &=& r^2-2Mr+a^2+\frac{Q^2 r}{M}\\
    \rho^2 &=& r^2 + a^2\cos^2\theta +\frac{Q^2 r}{M}
\end{eqnarray}
with $a<\sqrt{(M-Q^2/2M)^2}$. The radii of the event horizon $r_H$ and the stationary limit surface $r_g$ are given by
\begin{eqnarray}
    r_H &=& M-\frac{Q^2}{2 M}+\sqrt{\left(M-\frac{Q^2}{2 M}\right)^2-a^2}\\
    r_g(\theta)&=& M-\frac{Q^2}{2M}+\sqrt{\left(M-\frac{Q^2}{2 M}\right)^2-a^2 \cos^2\theta}
\end{eqnarray}
The inequality representing the photon region is deduced through the separated geodesic equations of radial and polar coordinates \cite{KN and KS BHs, Null Geodesics in KS} and is given as
\begin{eqnarray}
   \left( \eta_{KS}+a^2 \cos^2\theta\right) \sin^2\theta &\geq& \Phi_{KS}^2 \cos^2\theta
\end{eqnarray}
where
\begin{eqnarray}
    \eta_{KS} &=& -\frac{r^2\left[\left(2M^2\left(r^2-3Mr-Q^2\right)+3MQ^2 r+Q^4\right)^2-8a^2M^4\left(2Mr+Q^2\right)\right]}{a^2 M^2(2Mr-2M^2+Q^2)^2}\nonumber\\
    \Phi_{KS} &=& \frac{2a^2M^2(M+r)+a^2MQ^2-6M^3r^2-2M^2Q^2 r+2M^2r^3+3MQ^2r^2+Q^4 r}{aM(2M^2-2Mr-Q^2)}\nonumber
\end{eqnarray}
Using this inequality, the photon region, ergoregion, and the horizon are plotted below in Fig. \ref{Kerr-Sen illustration} for black hole parameters $\{M, a, Q\}=\{4,2,2\}$ and $\{4,3,2\}$. For the same set of black hole parameters, boundary curves $r_i(\theta)$ and $r_o(\theta)$ of the inner and outer boundaries are evaluated using the Eqs. (\ref{inner boundary equation}) and (\ref{outer boundary equation}), and are plotted along with the stationary limit surface in Fig. \ref{consistency of Kerr-Sen boundaries}. The curves $r_c(\theta)$ of the boundaries deduced from the separated radial and polar geodesic equations are also plotted in the same figure, which satisfy
\begin{eqnarray}
   \left( \eta_{KS}+a^2 \cos^2\theta\right) \sin^2\theta &=& \Phi_{KS}^2 \cos^2\theta
\end{eqnarray}
\begin{figure}[H]
\centering
\begin{subfigure}{0.40\textwidth}
\centering
\includegraphics[scale=0.40]{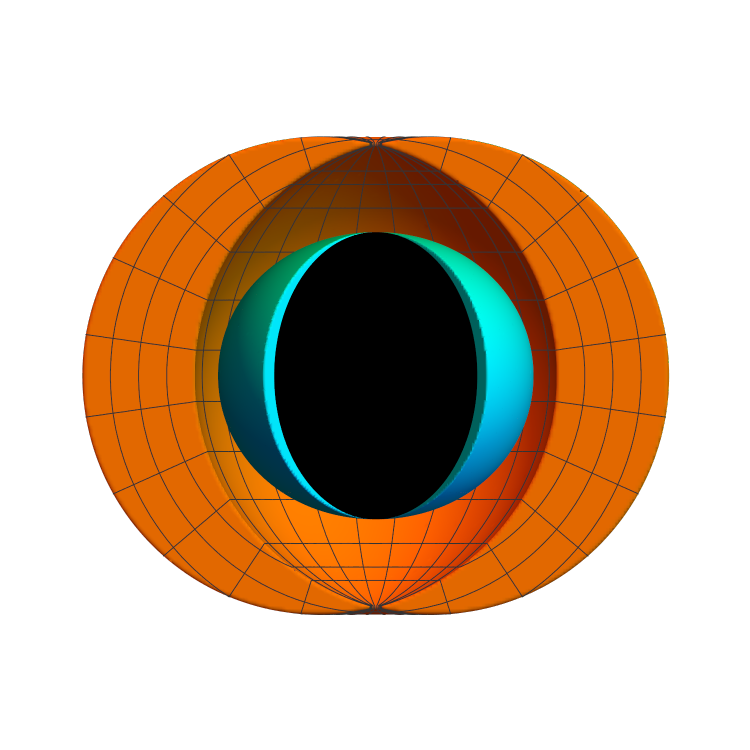}
\caption{$M=4$, $a=2$, $Q=2$}
\end{subfigure}
\hfill
\begin{subfigure}{0.40\textwidth}
\centering
\includegraphics[scale=0.40]{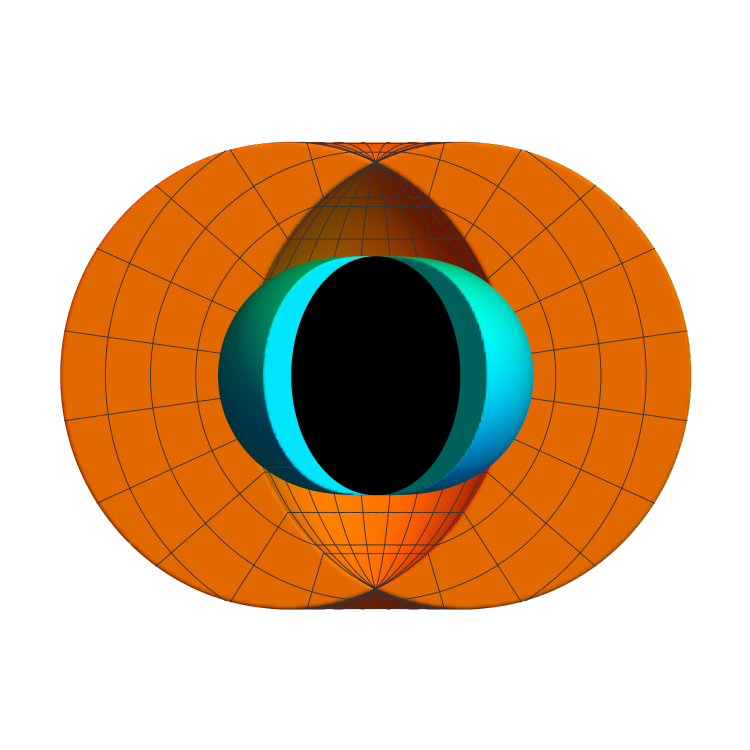}
\caption{$M=4$, $a=3$, $Q=2$}
\end{subfigure}
\caption{3D illustrations showing the photon region in orange, ergoregion in cyan, and the event horizon in black for Kerr-Sen black holes.}
    \label{Kerr-Sen illustration}
\end{figure}
From the figure, one can note that the curves $r_c(\theta)$ overlap with the curves $r_i(\theta)$ and $r_o(\theta)$, thus proving the consistency of the two approaches for the Kerr-Sen black hole.

\begin{figure}[H]
\centering
\begin{subfigure}{0.40\textwidth}
\centering
\includegraphics[scale=0.30]{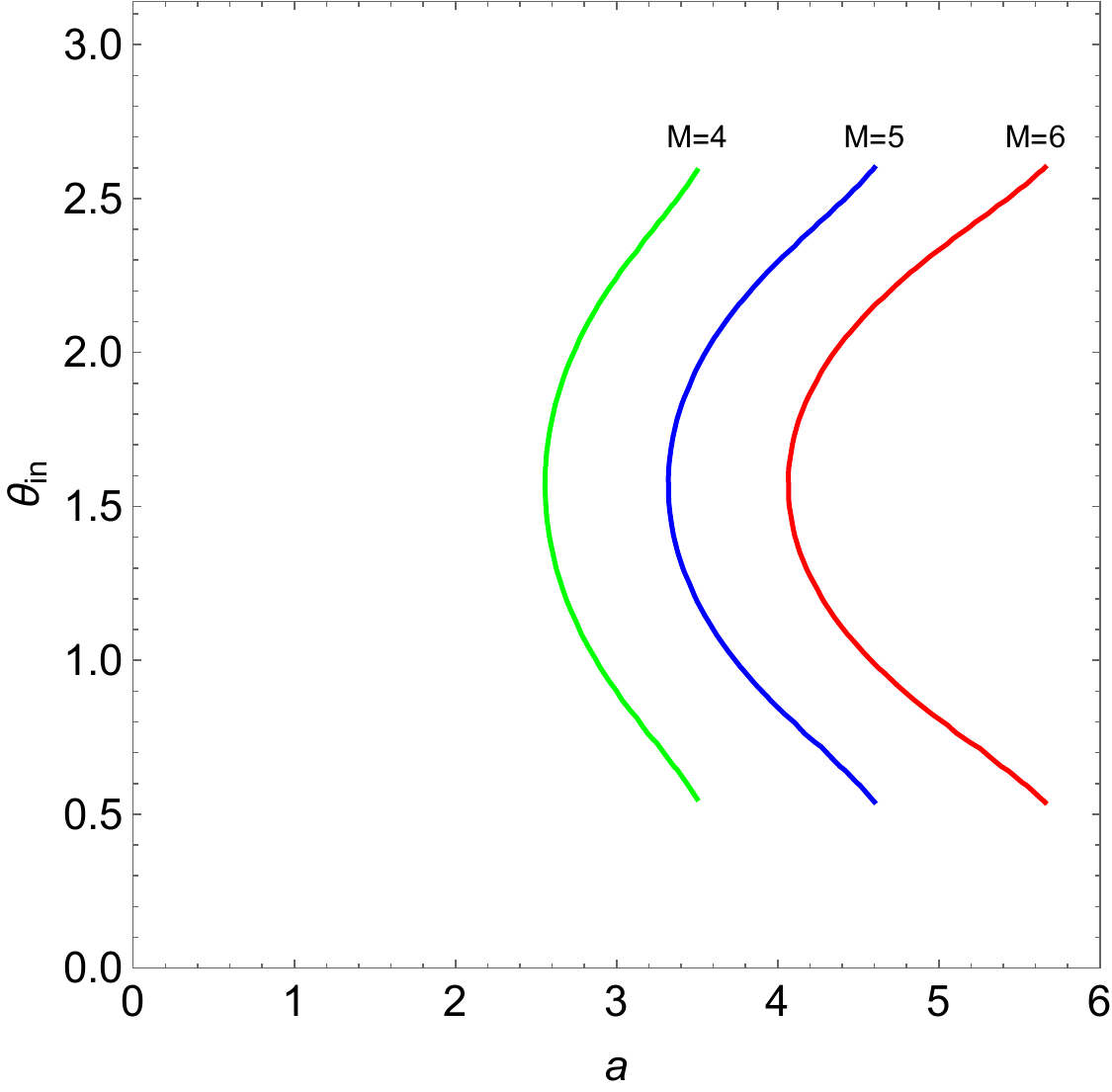}
\caption{$Q=2$}
\end{subfigure}
\hfill
\begin{subfigure}{0.40\textwidth}
\centering
\includegraphics[scale=0.30]{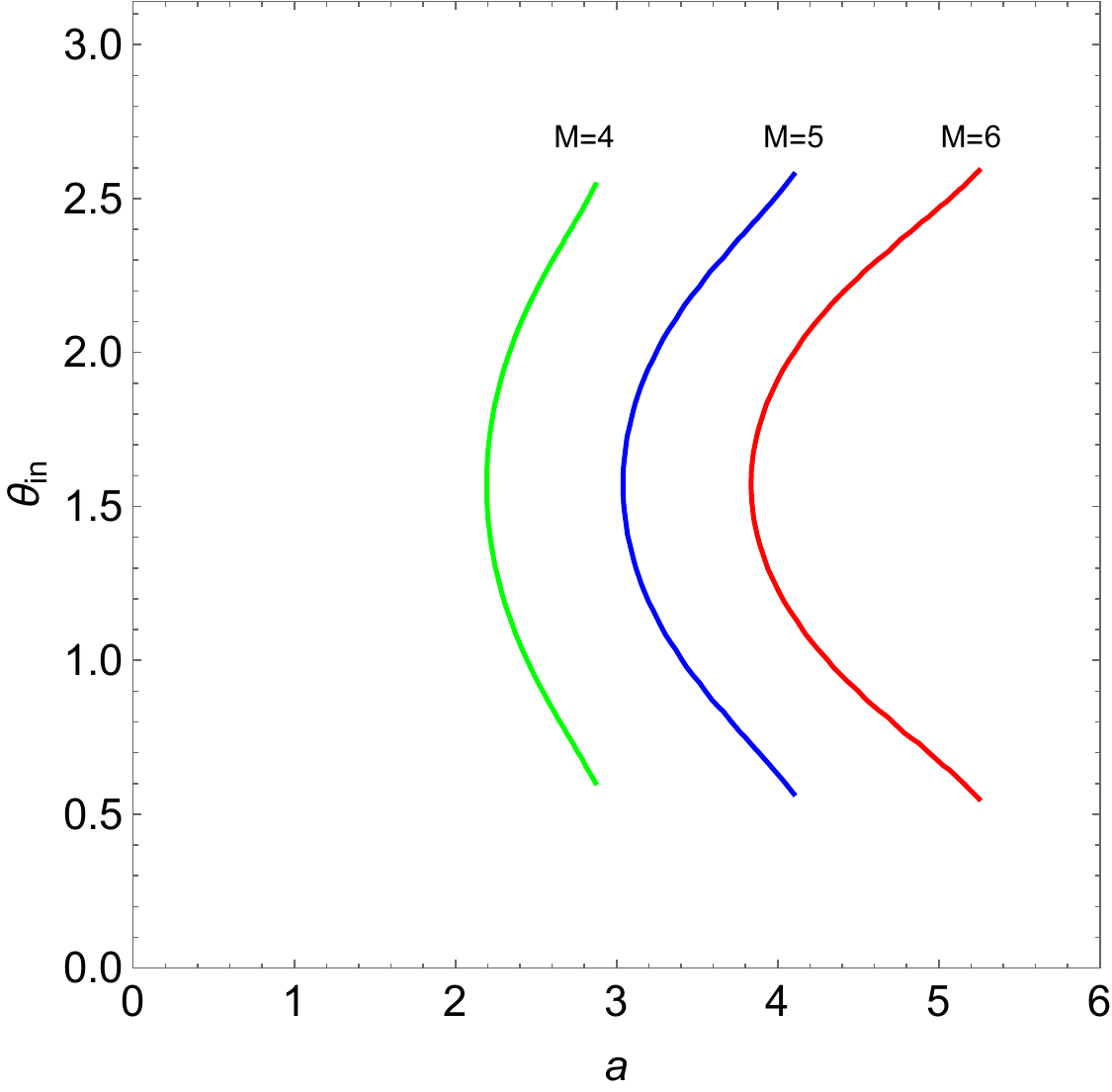}
\caption{$Q=3$}
\end{subfigure}
\caption{Intersection points $\theta_{in}$ for black hole mass $M=4,5,6$ and the spin parameter $a<\sqrt{(M-Q^2/2M)^2}$}\label{variation with spin in Kerr-Sen}
\end{figure}

The points of intersection $\theta_{in}$ of the inner boundary and the stationary limit surface are plotted using Eq.(\ref{Intersection}) for black holes of mass $M=\{4,5,6\}$ and charges $Q=\{2,3\}$ over the range $a<\sqrt{(M-Q^2/2M)^2}$ of the spin parameter in Fig. \ref{variation with spin in Kerr-Sen}.
The critical spin corresponding to the single intersection point in the equatorial plane is derived as a function of mass $M$ and charge $Q$ of the black hole and is simply written as,
\begin{eqnarray}
    a_c &=& \frac{2M^2-Q^2}{\sqrt{8M^2-2Q^2}}
\end{eqnarray}

\section{Discussion} \label{conclusion}

We provide a formal construction of the photon region boundaries for generic asymptotically flat, stationary axisymmetric black hole spacetimes admitting spherical photon orbits (SPOs). Explicit expressions are given to determine the location of the photon region boundary. Further, by analyzing the behavior of the metric functions near the Killing horizon and the asymptotic spatial infinity, we showed that the well-defined solutions for the boundary curves exist in the domain of outer communication. The results are valid in any metric theory of gravity since no information regarding the field equations is incorporated at any step of the investigation. 

For the analysis, we have considered that the background spacetime admits SPOs. However, in general, a stationary axisymmetric spacetime may or may not admit SPOs. In the spacetimes that don't admit SPOs, one may need to consider photon orbits that are not confined to $r=$constant hypersurfaces. Such a class of photon orbits called the fundamental photon orbits (FPOs) has been explored in the background of the Kerr black hole with proca hair \cite{FPOs}. However, it is not evident whether such FPOs lie in a closed region of spacetime and whether one can construct a photon region of such FPOs. In the current work, we have restricted our attention to the photon region of SPOs and the spacetimes that allow the existence of SPOs. It would be intriguing to extend the current understanding of the photon region to a region, if at all exist, where a set of FPOs are confined to lie. 

It has been argued that a stationary axisymmetric spacetime admitting SPOs may not necessarily have separable $r$ and $\theta$ components of geodesic equations \cite{photon trapping orbits}. For a spacetime admitting SPOs to be separable, the ratio of the metric components $g_{rr}$ and $g_{\theta\theta}$ needs to have a special form $g_{\theta\theta}/g_{rr}=f(r) h(\theta)$. In this work, since we have not considered any special form for the ratio $g_{\theta\theta}/g_{rr}$, our results are applicable to both separable and non-separable black hole spacetimes. Analogously, our construction does not rely on any constants of the photon's motion, and the photon region boundary is entirely derived from the generic form of the background metric functions. 

In case of non-separable known black hole metrics, wherein the analytical examination of SPOs and photon region may lead to complicated equations, our construction can provide the location of the photon region boundary as the initial condition for numerical estimation of SPOs and photon region. Such an implementation of our technique through numerical illustrations would also be interesting to follow.

Overall, our results can be used in the following scenarios: (i) In the case where it is known that SPOs exist (separable or non-separable), our method locates the boundary of the photon region directly from the metric functions without the need of the separation constant or a Carter-like constant. (ii) In the case where the existence of SPOs has not been ascertained, our method provides the potential locations where they may be found numerically. It provides a starting point or initial condition for numerical analysis. This could lead to any of the two conclusions: that SPOs may or may not exist in the chosen spacetime. (iii) Our method will also be useful to find the SPOs in a case where a spacetime possesses the symmetry required for separability (in principle), but the separate geodesic equations for $r$ and $\theta$ coordinates have not yet been deduced analytically, due to say, a complicated form of the Hamilton-Jacobi equation.

A subclass of the FPOs, called “spheroidal" orbits, is considered in \cite{photon trapping orbits}. These orbits are confined on a spheroidal-shaped shell and could intersect the equatorial plane at more than one radius while being bounded between two radii $r_1$ and $r_2$. It has been argued that a spheroidal orbit can be reduced to a SPO by an appropriate coordinate transformation, which may result in the spacetime metric with an additional $g_{r\theta}\neq 0$ component. Consequently, in some stationary axisymmetric geometries, the presence/absence of SPOs could be a result of a specific coordinate choice. If the coordinate transformation, say $(r,\theta)\to (r',\theta')$, reduces the spheroidal orbits to SPOs while conserving the metric form as in Eq.(\ref{metric}) with vanishing $g_{r\theta}$ component, our construction of the photon region boundaries can be implemented in the background of the transformed metric in $(r',\theta')$ coordinates.

Following this construct, we identified some interesting features of the photon region, namely: The outer boundary of the photon region is always situated outside the ergoregion and is populated only by the counter-rotating photon orbits. Whereas, the inner boundary is populated by co-rotating photon orbits. The photon region and the ergoregion overlap partially for a specific range of black hole parameters. In this range, the inner boundary of the photon region intersects the stationary limit surface. One can then expect a set of limiting values of black hole parameters beyond which there is no overlap between the two regions. The black hole geometries with the limiting parameter values (for example, the critical spin in section \ref{Examples}) are then likely to have only one common point between the photon region and the ergoregion in the $(r,\theta)$ plane.

Further, light rings are found to be located at the extrema along the polar angular direction of the curves $r_{i}(\theta)$ and $r_{o}(\theta)$ representing the inner and outer boundaries of the photon region, respectively. This finding supports the presence of at least two light rings: one co-rotating light ring on the inner boundary and one counter-rotating light ring on the outer boundary of the photon region, which is consistent with the results of \cite{Stationary BHs and LRs}.

For horizon-less ultra-compact objects, the light rings are shown to occur in pairs for each rotation sense \cite{LR Stability in UCOs}. Then, by extending the current understanding of light rings on the photon region boundaries to ultra-compact objects, one can check whether one gets multiple photon region boundaries in such spacetimes, which would be interesting to investigate further \cite{PR of UCO}.

\section*{Acknowledgment}
We thank Pedro V.P. Cunha for a helpful discussion on the fundamental photon orbits and the existence of spherical photon orbits and the anonymous referee for useful comments.

\end{document}